\documentclass[DIV=12,numbers=noenddot]{scrartcl}

\usepackage[utf8]{inputenc}
\usepackage{slashed}
\usepackage{booktabs}
\usepackage{multirow}
\usepackage{graphicx}

\usepackage[section]{placeins}
\usepackage{longtable}
\usepackage{tabu}
\usepackage{lmodern}
\usepackage{amsmath,amssymb}
\usepackage[usenames]{xcolor}
\usepackage{scalefnt}
\usepackage{xspace}
\usepackage{listings}
\usepackage{booktabs}
\usepackage{authblk}
\usepackage[]{units}
\usepackage[normalem]{ulem}

\usepackage{tikz}
\usetikzlibrary{arrows,shapes}
\usetikzlibrary{trees}
\usetikzlibrary{matrix,arrows} 				
\usetikzlibrary{positioning}				
\usetikzlibrary{calc,through}				
\usetikzlibrary{decorations.pathreplacing}  
\usepackage{pgffor}							
\usetikzlibrary{decorations.pathmorphing}	
\usetikzlibrary{decorations.markings}

\usepackage[font=small,labelfont=bf,format=plain,margin=0.05\textwidth]{caption}
\usepackage{subcaption}
\graphicspath{{Plots/}}
\numberwithin{equation}{section}

\newcommand{\sq}{{\tilde{q}}}
\newcommand{\gl}{{\tilde{g}}}
\newcommand{\Os}{O_{\mathrm{s}}}
\newcommand{\msq}{{m_{\tilde{q}}}}
\newcommand{\msqq}{{m^2_{\tilde{q}}}}
\newcommand{\msqqu}{{m^4_{\tilde{q}}}}
\newcommand{\mgl}{{m_{\tilde{g}}}}
\newcommand{\mglq}{{m^2_{\tilde{g}}}}
\newcommand{\mglqu}{{m^4_{\tilde{g}}}}
\newcommand{\mOs}{m_{O_{\mathrm{s}}}}

\newcommand{\als}{\alpha_{\mathrm{s}}}
\newcommand{\alssusy}{\hat{\alpha}_{\mathrm{s}}}
\newcommand{\myth}{\mathrm{th}}
\newcommand{\ii}{\mathrm{i}}
\newcommand{\SUNN}{N_c}
\newcommand{\mygamma}{\gamma_{\mathrm{E}}}
\newcommand*{\polylog}{\text{Li}_2}
\allowdisplaybreaks

\usepackage[
  pdftitle={Squark production with R-symmetry beyond NLO at the LHC},
  pdfauthor={C. Borschensky, F. Frisenna, W. Kotlarski, A. Kulesza, D. Stoeckinger},
  pdfkeywords={MRSSM},
  bookmarks=true, linktocpage, colorlinks=true, allbordercolors=white,
  allcolors=blue]{hyperref}
\usepackage{orcidlink} 

\usepackage[
  backend=bibtex,
  bibencoding=utf8,
  sorting=none,
  url=true,
  doi=false,
  style=phys,
  biblabel=brackets,
  subentry=false,
  eprint=true,
  maxbibnames=5
]{biblatex}
\titlehead{
\begin{flushright}
	KA-TP-01-2024\\
	MS-TP-24-05
\end{flushright}
}

\title{\boldmath Squark production with R-symmetry \\ beyond NLO at the LHC}
\date{}
\author[1]{Christoph Borschensky\orcidlink{0000-0002-8486-8784}\thanks{\href{mailto:christoph.borschensky@kit.edu}{christoph.borschensky@kit.edu}}}
\author[2]{Fausto Frisenna\thanks{\href{mailto:fausto.frisenna@uni-muenster.de}{fausto.frisenna@uni-muenster.de}}}
\author[3]{Wojciech Kotlarski\orcidlink{0000-0002-1191-6343}\thanks{\href{mailto:wojciech.kotlarski@ncbj.gov.pl}{wojciech.kotlarski@ncbj.gov.pl}}}
\author[2]{Anna Kulesza\thanks{\href{mailto:anna.kulesza@uni-muenster.de}{anna.kulesza@uni-muenster.de}}}
\author[4]{Dominik St\"ockinger\thanks{\href{mailto:dominik.stoeckinger@tu-dresden.de}{dominik.stoeckinger@tu-dresden.de}}}

\affil[1]{Institute for Theoretical Physics, Karlsruhe Institute of Technology, Wolfgang-Gaede-Str. 1, 76131 Karlsruhe, Germany}
\affil[2]{Institute for Theoretical Physics, University of M\"unster, Wilhelm-Klemm-Straße 9, 48149 M\"unster, Germany}
\affil[3]{National Centre for Nuclear Research, Pasteura 7, 02-093 Warsaw, Poland}
\affil[4]{Institut f\"ur Kern- und Teilchenphysik, Zellescher Weg 19, 01069 Dresden, Germany}

\begin{document}
\maketitle
\begin{abstract}
The Minimal R-symmetric Supersymmetric Standard Model (MRSSM) provides a realisation of supersymmetry in which the parameter space is less constrained by the current LHC data than in the simplest supersymmetric scenarios. In the present paper, we obtain the most precise theoretical predictions in the MRSSM for squark production at the LHC, enabling accurate interpretations of LHC data in terms of the MRSSM.
We perform threshold resummation of soft gluon corrections to the total cross sections for the production of squark-(anti)squark pairs at the LHC in the MRSSM framework. The resummation is carried out using the direct QCD method and reaches the next-to-next-to-leading-logarithmic (NNLL) accuracy, which requires calculating the one-loop matching coefficients in the relevant production channels. The resummed cross sections are then matched to the available NLO results and evaluated for $\sqrt{S}=13.6$ TeV. 
Compared with the Minimal Supersymmetric Standard Model (MSSM), the cross sections in the MRSSM can be significantly reduced, implying less stringent limits on squark and gluino masses. Our results carry significant implications for exploring the viability of supersymmetry at the LHC. 
The results of our calculation are publicly available as a numerical package.
\clearpage
\end{abstract}

\tableofcontents

\section{Introduction}
The search for new particles is one of the primary tasks of the experiments at the Large Hadron Collider (LHC). Supersymmetry (SUSY) \cite{Wess:1973kz, Wess:1974tw, Fayet:1976et, Farrar:1978xj, Sohnius:1985qm,Martin:1997ns} remains one of the best-motivated ideas for physics beyond the Standard Model. SUSY predicts a plethora of new particles which --- if their masses are around the TeV scale --- can ameliorate the electroweak hierarchy problem and contribute to the dark matter in the universe. So far, LHC experiments have not observed signals for such SUSY particles, but it should be noted that SUSY can be realised in several distinct ways with different properties.

The most common SUSY model is the Minimal Supersymmetric Standard Model (MSSM). An alternative is provided by the Minimal R-symmetric Supersymmetric Standard Model (MRSSM). Both are minimal but in different respects. The usual MSSM predicts the minimal particle content: besides the Standard Model (SM), the MSSM introduces a second Higgs doublet and SUSY partners to all particles. Its strongly interacting sector contains squarks $\tilde{q}_L$, $\tilde{q}_R$ corresponding to all quark flavours and chiralities and a Majorana gluino. All up-type squarks and all down-type squarks can mix, leading to potentially large flavour-changing neutral current interactions and the SUSY flavour problem.

The MRSSM, in contrast, has a higher degree of symmetry and, as a result, predicts more particles but has fewer free parameters~\cite{Kribs:2007ac}. It postulates a global U(1) R-symmetry~\cite{Fayet:1974pd, Salam:1974xa} under which SM states are uncharged, while SUSY particles such as squarks and gluinos are charged. The R-charges of $\tilde{q}_L$ and $\tilde{q}_R$ are opposite, such that left-right squark mixing is forbidden and many free parameters related to flavour mixing are absent in the MRSSM. The nonzero R-charge of the gluino implies that the gluino is a Dirac instead of a Majorana fermion. All of these differences strongly impact the LHC phenomenology of the MRSSM.

One important consequence is that several squark production modes are forbidden in the MRSSM. Specifically, the production of $\tilde{q}_L\tilde{q}_L$ pairs or $\tilde{q}^{\*}_L\tilde{q}^{*}_R$ pairs (or similar pairs with $R\leftrightarrow L$ exchanged) is forbidden by R-charge conservation in the MRSSM. Another important difference affects the production of e.g. $\tilde{q}_L\tilde{q}_R$ or $\tilde{q}^{\*}_L\tilde{q}^*_L$ pairs. These production channels are allowed both in the MSSM and the MRSSM, but due to the Majorana nature of the gluino in the MSSM there is only a $1/m_{\tilde{g}}$ suppression of gluino $t$-channel diagrams for heavy gluino masses $m_{\tilde{g}}$, while the Dirac gluino contribution in the MRSSM is more strongly suppressed as $1/m_{\tilde{g}}^2$.

Because of these differences, the LHC production cross sections for squarks are significantly suppressed compared to the MSSM. This suppression is particularly important since for squark and gluino masses within LHC reach, squark production processes typically dominate. The phenomenological consequences have been quantitatively explored in
Refs.~\cite{Heikinheimo:2011fk, Kribs:2012gx, Kribs:2013oda, Diessner:2017ske, Diessner:2019bwv} with increasing accuracy. The result is that in the MRSSM, smaller squark masses are more compatible with the negative LHC search results than in the MSSM. Specifically, Ref.~\cite{Diessner:2019bwv} found approximately 600 GeV lower squark mass limits in the MRSSM compared to the MSSM, based on a next-to-leading order theory calculation.

The MRSSM has several further attractive features beyond increasing the viable parameter space of squark masses. The absence of left-right squark mixing eliminates many parameters responsible for flavour-violating interactions and thus alleviates the SUSY flavour problem. The electroweak and Higgs sectors contain new interactions which can push the Higgs boson mass up to the observed value for smaller values of top-squark masses than in the MSSM~\cite{Bertuzzo:2014bwa, Diessner:2014ksa, Diessner:2015yna, Diessner:2015iln}, and which can contribute to the $W$-boson mass~\cite{Diessner:2014ksa, Diessner:2015iln, Athron:2022isz}.
The MRSSM also contains various possibilities to explain dark matter~\cite{Belanger:2009wf, Chun:2009zx, Buckley:2013sca}, particularly in connection with an extra light singlet Higgs~\cite{Diessner:2015iln} as well as colour-octet scalars \cite{Choi:2008ub,Plehn:2008ae,Goncalves-Netto:2012gvn,Kotlarski:2016zhv,Darme:2018dvz} and Dirac gauginos \cite{Choi:2010gc,Choi:2009ue,Chalons:2018gez}.
The special flavour physics properties have also been studied in the lepton and top sectors in Refs.~\cite{Dudas:2013gga, Fok:2010vk, Herquet:2010ka} and including muon $\mathit{g}\!-\!2$ in Ref.~\cite{Kotlarski:2019muo}.

Despite the strong motivation of the MRSSM, so far, the theoretical predictions for LHC processes have not reached the same level of precision as in the MSSM, with the corresponding calculations reported in Refs.~\cite{Beenakker:1994an, Beenakker:1996ch, Kulesza:2008jb, Kulesza:2009kq, Beenakker:2009ha, Beneke:2009rj, Beenakker:2010nq, Beneke:2010da, Beenakker:2011fu, Beenakker:2011sf, Falgari:2012hx, Langenfeld:2012ti,  Goncalves-Netto:2012nvl, Pfoh:2013iia, Beenakker:2013mva, Gavin:2013kga, Beneke:2013opa, Broggio:2013cia, Beenakker:2014sma, Gavin:2014yga, Beneke:2014wda, Beenakker:2016gmf, Beenakker:2016lwe, Beneke:2016kvz}. The present paper is dedicated to an improved evaluation of the squark production cross section in the MRSSM. It includes fixed-order next-to-leading order (NLO) corrections (as already discussed in Ref.~\cite{Diessner:2017ske}) and the resummation of large logarithmic corrections due to soft gluon emission close to the production threshold.

Including the soft-gluon resummation is essential because the MRSSM squark cross sections are suppressed, but resummation is expected to increase the cross sections, with the effect most pronounced for heavy squark masses. Carrying out threshold resummation requires process-specific information on the LO contributions and higher-order non-logarithmic terms which do not vanish at threshold, all decomposed in a chosen colour basis. As we aim here to perform next-to-next-to-leading logarithmic (NNLL) resummation for the colour channels which in the threshold limit receive the dominant contribution in the $s$-wave configuration, we need to calculate, in addition to the LO cross sections, one-loop virtual corrections in the MRSSM for the relevant colour channels. For the channels which are $p$-wave dominated at threshold, i.e.\ suppressed, the resummation is carried out at the next-to-leading logarithmic (NLL) accuracy, in accordance with the approach adopted in the MSSM case~\cite{Beenakker:2014sma, Beenakker:2016gmf, Beenakker:2016lwe}.
The computation presented here will allow future LHC analyses to be interpreted both in the MSSM and the MRSSM on an equal footing and to obtain precise information on the additional squark mass range viable in the MRSSM.

The paper is organised as follows. In Sec.~\ref{sec:MRSSM} we collect the relevant details of the MRSSM and the squark cross sections at leading order. Sec.~\ref{sec:resummation} describes the resummation of soft-gluon contributions in the MRSSM and the combination with fixed-order NLO corrections. Sec.\ \ref{sec:results} presents numerical results for the cross sections, the differences between the MRSSM and the MSSM and for the remaining uncertainties, and Sec.\ \ref{sec:conclusions} presents conclusions. The Appendix provides the leading-order colour-split cross sections and the analytical results for the one-loop hard-matching coefficient relevant for threshold resummation.

\section{The Minimal R-symmetric Supersymmetric Standard Model}\label{sec:MRSSM}
In this section, we provide details on the MRSSM and its strongly interacting sector. We also define the input parameters and the employed renormalisation scheme.

\subsection{Field content, Lagrangian and renormalisation scheme}
The defining property of the MRSSM is the unbroken continuous U(1) R-symmetry under which SM fields are uncharged. R-symmetry does not commute with SUSY, and in a superspace formulation, the Grassmann coordinates $\theta$ transform into $e^{\ii\alpha}\theta$. The resulting nonvanishing R-charges of superpartners to SM fields depend on the structure of the supermultiplet.

Left-handed squarks $\tilde{q}_L$ are part of chiral supermultiplets, and the resulting R-charge of $\tilde{q}_L$ is $R=+1$. In contrast, right-handed squarks $\tilde{q}_R$ are part of antichiral supermultiplets and have opposite R-charge $R=-1$. Gluinos $\tilde{g}$ have R-charge $R=+1$ and are Dirac fermions. The left-handed gluino and the right-handed antigluino behave like their MSSM counterparts. The additional Dirac gluino degrees of freedom are absent in the MSSM; they are described by an additional chiral supermultiplet which does not admit direct couplings to quarks or squarks. However, they have their own SUSY partners, which are spin-0 colour octets $O^a$, the so-called scalar gluons (sgluons) with $R = 0$.

We illustrate the field and particle content by reproducing the Lagrangian for soft SUSY breaking in the coloured sector; see also Ref.~\cite{Diessner:2017ske}:
\begin{align}
\mathcal{L}_{\mathrm{soft}} =\ & -\frac{(m_{\tilde{q}_L}^2)_{ij}}{2}
\tilde{q}_{iL}^{*}\tilde{q}_{jL}^{\*} -
\frac{(m_{\tilde{q}_R}^2)_{ij}}{2}
\tilde{q}_{iR}^{*}\tilde{q}_{jR}^{\*}-m_{O}^2\left|O^{a}\right|^2 -
\mgl\overline{\tilde{g}}\tilde{g} +\mgl\left(\sqrt{2}D^a
O^a + \text{h.c.}\right)\;.
\end{align}
Here, $i$ and $j$ are flavour indices running over the six flavours of up- and down-type quarks. The squark mass matrices can have significant off-diagonal components in flavour space. As mentioned in the Introduction, other flavour-violating terms present in the MSSM are absent here, alleviating the flavour problem of the MRSSM. The $D^a$ are auxiliary fields of the $\mathrm{SU(3)}$ vector SUSY multiplet, which can be eliminated in favour of squark fields, leading to interactions between sgluons and squarks which influence LHC squark production at the one-loop level. The sgluon field can be decomposed as $O=({O_s +\ii O_p})/{\sqrt{2}}$, such that the tree-level masses of the pseudoscalar $O_p$ and the scalar $O_s$ are given by $m_{O_p}=m_{O}$ and $m_{O_s} = \sqrt{m_{O}^2+ 4 \mglq}$, respectively.

In the calculations presented in this paper, we assume the absence of squark flavour-violating parameters and, thus, the absence of flavour mixing. The free parameters in the coloured sector are then the masses $\msq$ for each squark flavour and each chirality, the gluino mass $\mgl$, the sgluon mass parameter $m_O$ and the strong coupling constant $\als$.

\tikzset{
    vector/.style={decorate, decoration={snake}, draw},
	provector/.style={decorate, decoration={snake,amplitude=2.5pt}, draw},
	antivector/.style={decorate, decoration={snake,amplitude=-2.5pt}, draw},
    fermion/.style={draw=black, postaction={decorate},
        decoration={markings,mark=at position .55 with {\arrow[draw=black]{>}}}},   
    fermionbar/.style={draw=black, postaction={decorate},
        decoration={markings,mark=at position .55 with {\arrow[draw=black]{<}}}},
    fermionnoarrow/.style={draw=black},
    gluon/.style={decorate, draw=black,
        decoration={coil,aspect=0.5,amplitude=4pt, segment length=4pt}},
    scalar/.style={dashed,draw=black, postaction={decorate},
        decoration={markings,mark=at position .55 with {\arrow[draw=black]{>}}}},
    scalarbar/.style={dashed,draw=black, postaction={decorate},
        decoration={markings,mark=at position .55 with {\arrow[draw=black]{<}}}},
    scalarnoarrow/.style={dashed,draw=black},
    electron/.style={draw=black, postaction={decorate},
        decoration={markings,mark=at position .55 with {\arrow[draw=black]{>}}}},
	bigvector/.style={decorate, decoration={snake,amplitude=4pt}, draw},
	ghost/.style={dotted,draw=black, postaction={decorate},
        decoration={markings,mark=at position .55 with {\arrow[draw=black]{>}}}},
	ghostbar/.style={dotted,draw=black, postaction={decorate},
        decoration={markings,mark=at position .55 with {\arrow[draw=black]{<}}}},
         gluonloop/.style={decorate, draw=black,
        decoration={coil,aspect=0.6,amplitude=3pt, segment length=4pt}},   
}

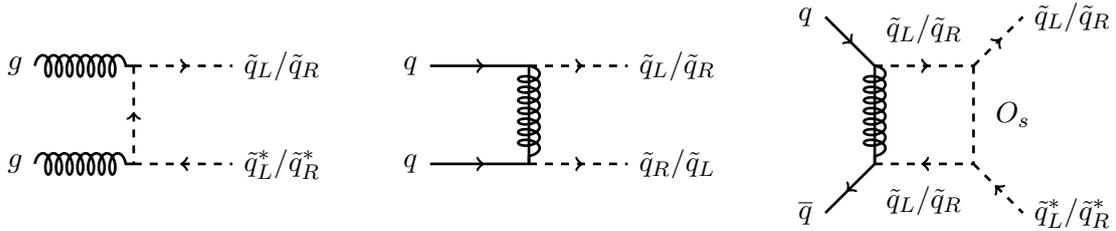
\begin{figure}[tp]
\begin{center}
\begin{tikzpicture}[line width=1.0 pt, scale=1.3, arrow/.style={thick,->,shorten >=2pt,shorten <=2pt,>=stealth}]

\begin{scope}[shift={(0,0)}]
	\draw[gluon] (-1,0.5)--(0,0.5);
	\draw[gluon] (-1,-0.5)--(0,-0.5);
	\node at (-1.2,0.5){$g$};
	\node at (-1.2,-0.5){$g$};
	\draw[scalarbar] (0,0.5)--(0,-0.5);
	\draw[scalar] (0,0.5)--(1,0.5);
	\draw[scalarbar] (0,-0.5)--(1,-0.5);
	\node at (1.5,0.5){$\sq_L/\sq_R$};
	\node at (1.5,-0.5) {$\sq^{*}_L/\sq_R^{*}$};
\end{scope}

\begin{scope}[shift={(4,0)}]
	\draw[fermion] (-1,0.5)--(0,0.5);
	\draw[fermion] (-1,-0.5)--(0,-0.5);
	\node at (-1.2,0.5){$q$};
	\node at (-1.2,-0.5){$q$};
	\draw[gluon] (0,0.5)--(0,-0.5);
	\draw[fermionnoarrow] (0,0.5)--(0,-0.5);
	\draw[scalar] (0,0.5)--(1,0.5);
	\draw[scalar] (0,-0.5)--(1,-0.5);
	\node at (1.5,0.5) {$\sq_L/\sq_R$};
	\node at (1.5,-0.5) {$\sq_R/\sq_L$};
\end{scope}

\begin{scope}[shift={(8,0)}]
	\draw[fermion] (-1,1)--(-0.5,0.5);
	\draw[fermionbar] (-1,-1)--(-0.5,-0.5);
	\node at (-1.2,1) {$q$};
	\node at (-1.2,-1) {$\overline{q}$};
	\draw[scalar] (0.5,0.5) -- (1,1);
	\draw[scalarbar] (0.5,-0.5) -- (1,-1);
	\node at (1.5,1) {$\sq_L/\sq_R$};
	\node at (1.5,-1) {$\sq^{*}_L/\sq_R^{*}$};
	\draw[scalar] (-0.5,0.5)--node[label=above:$\sq_L/\sq_R$]{}(0.5,0.5);
	\draw[scalarbar] (-0.5,-0.5)--node[label=below:$\sq_L/\sq_R$]{}(0.5,-0.5);
	\draw[scalarnoarrow] (0.5,0.5)--node[label=right:$O_s$]{}(0.5,-0.5);
	\draw[gluon] (-0.5,0.5)--(-0.5,-0.5);
	\draw[fermionnoarrow] (-0.5,0.5)--(-0.5,-0.5);
\end{scope}

\end{tikzpicture}
\end{center}
\caption{
Examples of tree-level and one-loop Feynman diagrams for the production of
squark-antisquark pairs and squark-squark pairs. The second diagram
exemplifies $t$-channel gluino exchange, where in the
MRSSM only the two indicated chirality combinations of squarks are possible because
of R-charge conservation.  The third diagram
exemplifies the appearance of the sgluon at the one-loop
level.
}\label{figwithtwotreelevelandoneloopdiagramsinvolvingsgluon}
\end{figure}

As another illustration, we reproduce sample Feynman diagrams relevant to squark production at LHC in Fig.~\ref{figwithtwotreelevelandoneloopdiagramsinvolvingsgluon}. The first diagram contributes to the production of squark-antisquark pairs at the tree-level, and the second diagram contributes to squark-squark production via $t$-channel gluino exchange. In the MRSSM, only the two indicated chirality combinations of squarks are possible because of R-charge conservation, and only the left-handed part of the gluino can contribute to the $t$-channel. In the MSSM, other chirality combinations are possible, and the Majorana mass term of the gluino can appear. For these reasons, the squark production rate is significantly lower in the MRSSM than in the MSSM. The third diagram exemplifies the appearance of the sgluon at the one-loop level. Together with renormalisation, loop diagrams with sgluons lead to non-decoupling, super-oblique corrections for heavy sgluons; see the discussion in Ref.~\cite{Diessner:2019bwv}.

Finally, we briefly define the renormalisation scheme employed here, and already in Ref.~\cite{Diessner:2017ske}, for the NLO computation. The relevant parameters were listed above and comprise the masses $\msq$, $\mgl$ and $m_O$ and the strong coupling constant $\als$. The masses of squarks and gluinos are defined in the on-shell scheme such that the parameters $\msq$ and $\mgl$ coincide with the respective physical pole masses. The sgluon mass needs no renormalisation since it only appears inside one-loop diagrams. The strong coupling is defined in a decoupling scheme such that the value of $\als$ entering the computation coincides with the well-known SM 5-flavour scheme used
in parton distribution functions. 

\subsection{Qualitative features and leading-order results}\label{sec:LOresults}
\begin{figure}[tp]
\begin{center}
\hspace{-0.15cm}
 \begin{subfigure}{0.33\textwidth}
 \includegraphics[width=\linewidth]{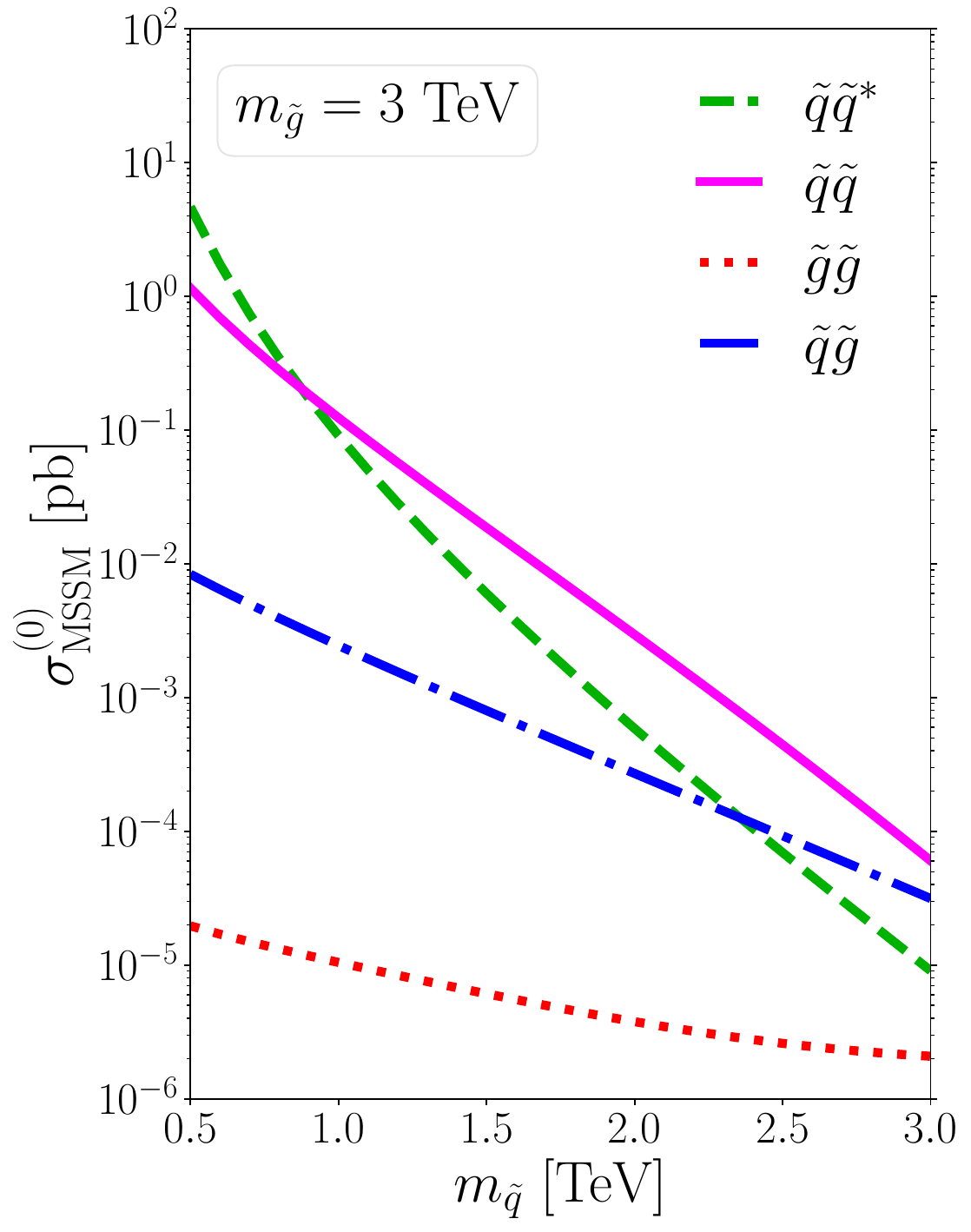}
 \end{subfigure}
 \hspace{-0.15cm}
 \begin{subfigure}{0.33\textwidth}
 \includegraphics[width=\linewidth]{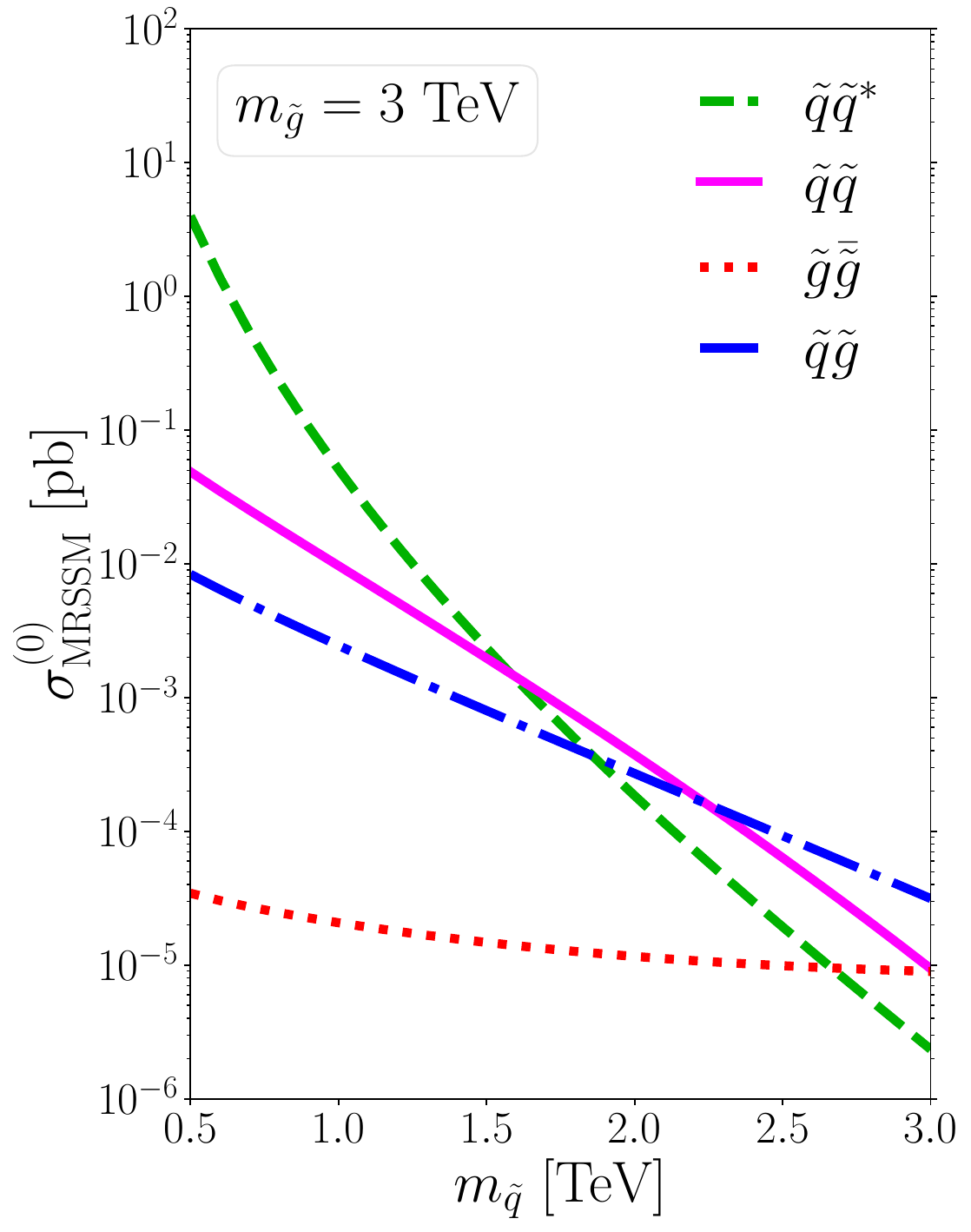}
 \end{subfigure}
  \hspace{-0.15cm}
 \begin{subfigure}{0.33\textwidth}
 \includegraphics[width=\linewidth]{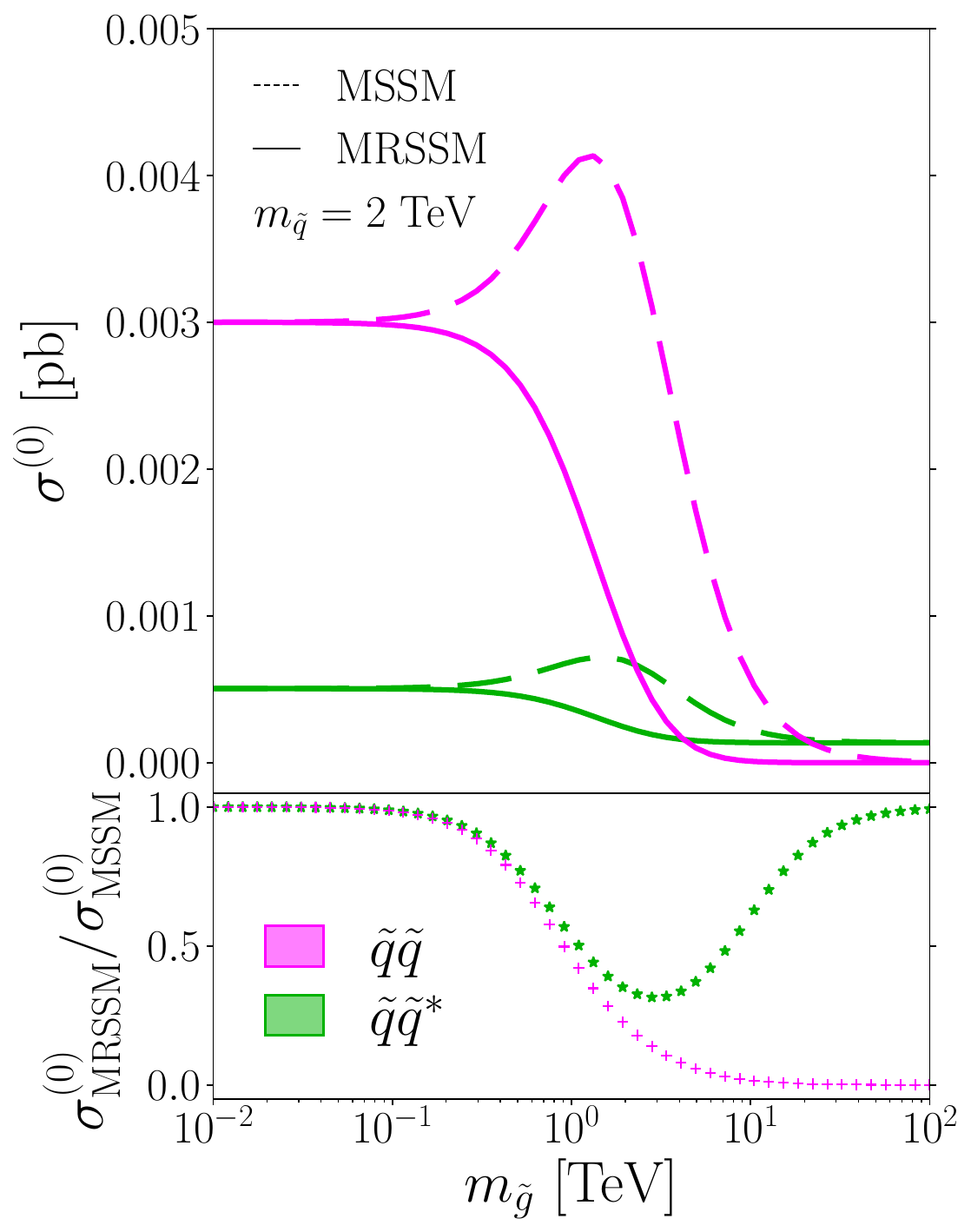}
 \end{subfigure}
\caption{LO cross sections at the LHC for $\sqrt{S} = \unit[13.6]{TeV}$, summed over the first five flavours. We used the Hessian PDF set of $\mathtt{PDF4LHC21}$~\cite{PDF4LHCWorkingGroup:2022cjn}, and set the scales to the average final state mass $\mu_R = \mu_F = \frac{m_1 + m_2}{2}$. Left plot: gluino and squark production in the MSSM at fixed $\mgl=3$ TeV. Center: equivalent processes in the MRSSM. Right: squark production across a wide gluino mass range, having fixed $\msq=2$ TeV.}
\label{fig: LOresultsat136}
\end{center}
\end{figure}

To further illustrate the differences between the MRSSM and the MSSM in Fig.~\ref{fig: LOresultsat136}, we show the LHC production cross sections of squarks and gluinos predicted at leading order (LO) by both models. Confronting the left and central plots, it is clear that the MRSSM exhibits extreme suppression for the squark pair production and a strong one for squark-antisquark production compared to the MSSM.
In addition, the dependence on the masses is different in the two models. The plots also show that, compared to the MSSM, the gluino production is enhanced in the MRSSM because of the additional degrees of freedom. However, this enhancement is less relevant in the parameter regions with a heavy gluino, which are theoretically preferred~\cite{Kribs:2007ac}. The right plot in Fig.~\ref{fig: LOresultsat136} portrays the differences between the MRSSM and MSSM differently and varies the gluino mass over a vast range while keeping a fixed squark mass. The phenomenologically most interesting region is the one where the squark and gluino masses are of a similar order, and the plot clearly shows the reduced MRSSM cross section in this region. However, the plot also highlights two interesting limits. First, in the (theoretical) limit of very light gluino masses, the difference resulting from the Dirac versus Majorana mass becomes
irrelevant; hence the MRSSM and MSSM cross sections become equal. Second, in the limit of ultra-heavy gluino masses, the gluino $t$-channel diagrams become negligible. Therefore in both models the squark-antisquark cross section effectively receives contributions from an equivalent subset of Feynman diagrams. Moreover, one can notice that in this decoupling limit, as expected by both models, the squark-squark production cross section approaches zero, but with a different power of the gluino mass, i.e.\ in the MRSSM the suppression is $\propto 1/\mgl^4$, while in the MSSM it is $\propto 1/\mglq$.

The LO cross sections and their behaviour have been discussed in detail in Ref.~\cite{Diessner:2017ske}; see also Refs.~\cite{Kribs:2012gx, Kribs:2013oda}. 
In App.~\ref{appendix: LO colour split} we provide analytical expressions for the LO cross sections decomposed in the $s$-channel colour basis, needed for the intended threshold resummation, cf.\ Sec.~\ref{sec:resummation}. This type of resummation enables to control large logarithmic contributions arising close the production threshold, i.e.\ in the limit $\sqrt{ \hat{s}} \to 2 \msq$, or, expressed in terms of the squark velocity $\beta\equiv\sqrt{1 - 4\msqq/\hat{s}}$, the limit $\beta \to 0$. In App.~\ref{appendix: LO colour split} we additionally provide the leading terms in the expansions of the colour-decomposed LO cross sections in $\beta$.

While all the expressions which can be found in App.~\ref{appendix: LO colour split} are given for the general $\mathrm{SU(\SUNN)}$ gauge group, here we present the leading terms in the threshold expansion for the colour-decomposed LO cross section specifically for $\SUNN=3$, with the colour basis indices appropriately adapted to the respective $\mathrm{SU(3)}$ representations.
For the $q_i q_j \to \sq \sq^{*}$ process, we obtain:
\begin{align}
\hat{\sigma}^{(0,\myth)}_{q_i \bar{q}_j \to \sq \sq^{*}, \mathbf{1}}&=\dfrac{64\pi\als^2}{243}\left[\beta^3\dfrac{\msqq}{\left(\mglq+\msqq\right)^2}\right], \\
\hat{\sigma}^{(0,\myth)}_{q_i \bar{q}_j \to \sq \sq^{*}, \mathbf{8}}&=\delta_{ij}\left\{\dfrac{\left(n_f-1\right)}{27\msqq}\pi\als^2\beta^3+\dfrac{4\pi \als \alssusy}{81}\left[\dfrac{\beta^3}{\left(\mglq+\msqq\right)}\right]\right\}+\dfrac{8\pi\alssusy^2}{243}\left[\beta^3\dfrac{\msqq}{\left(\mglq+\msqq\right)^2}\right].
\end{align}
The leading behavior of both singlet $\mathbf{1}$ and octet $\mathbf{8}$ cross sections is thus given by terms of ${\cal O}(\beta^3)$, i.e.\ the $p$-wave dominates, and there is no $s$-wave contribution. This constitutes a big difference to the MSSM as the R-symmetry forbids the production of different ``chiralities'' resulting in the suppression at the threshold. Note that in these and the following formulas, we distinguish the strong gauge coupling for gluon interactions $\als$ from the gluino-quark-squark coupling $\alssusy$, even though numerically $\als=\alssusy$ throughout (at higher orders the counterterm to both couplings differ in regularisation schemes which do not manifestly preserve supersymmetry, see Ref.~\cite{Diessner:2017ske} for details for the present case).

The leading terms in the expansion in $\beta$ of the $g g \to \sq \sq^{*}$ LO cross section split into colour channels read:
\begin{align}
\hat{\sigma}^{(0,\myth)}_{g g \to \sq \sq^{*}, \mathbf{1}}&=\dfrac{\left(n_f-1\right)\pi\als^2}{96}\left[\dfrac{\beta}{\msqq}\right], \\
\hat{\sigma}^{(0,\myth)}_{g g \to \sq \sq^{*}, \mathbf{8_S}}&=\dfrac{5\left(n_f-1\right)\pi\als^2}{192}\left[\dfrac{\beta}{\msqq}\right],\\
\hat{\sigma}^{(0,\myth)}_{g g \to \sq \sq^{*}, \mathbf{8_A}}&=\dfrac{\left(n_f-1\right)\pi\als^2}{64}\left[\dfrac{\beta^3}{\msqq}\right].
\end{align}
Importantly, near threshold the singlet $\mathbf{1}$ and the symmetric octet $\mathbf{8_S}$ squark-antisquark configurations are produced in the $s$-wave, while the antisymmetric octet $\mathbf{8_A}$ is suppressed by $\beta^2$. The results for $g g \to \sq \sq^{*}$ are the same as for the MSSM. This is to be expected as neither gluinos nor sgluons play a role at the lowest order.

Expanding the LO cross section for the $\sq\sq$ production near threshold yields the dominant terms:
\begin{align}
\hat{\sigma}^{(0,\myth)}_{q_i q_j \to \sq \sq, \bar{\mathbf{{3}}}}&=\dfrac{16\pi\alssusy^2}{81}\left[\beta^3\dfrac{\msqq}{\left(\mglq+\msqq\right)^2}\right], \\
\hat{\sigma}^{(0,\myth)}_{q_i q_j \to \sq \sq, \mathbf{6}}&=\dfrac{8\pi\alssusy^2}{81}\left[\beta^3\dfrac{\msqq}{\left(\mglq+\msqq\right)^2}\right].
\end{align}
Also in this case, the leading behavior is given by terms of ${\cal O}(\beta^3)$, and the production of the $\sq\sq$ pair proceeds in the $p$-wave. This power-suppressed behavior is different from the MSSM. It arises in the MRSSM since the chirality projectors effectively remove the gluino mass in the numerator of the gluino propagator in the expressions for $t$-channel diagrams.

\section{Threshold resummation}\label{sec:resummation}
In the case of production of a heavy-mass system, a significant contribution to the cross section comes from the region near threshold, where the partonic centre-of-mass-energy is close to the kinematic restriction for the on-shell production, here $\hat s \geq 4 \msqq$. The dominant corrections in this region arise either from soft-gluon emission off the initial- or final-state legs, or from an exchange of gluons between slowly moving coloured particles in the final state (Coulomb corrections).\footnote{The Coulomb corrections can be resummed as well~\cite{Beneke:2009rj, Beneke:2010da}, however such resummation is beyond the scope of this paper.}
The contributions due to soft-gluon radiation have the general form
\[
\als^n \log^m (\beta^2), \qquad \qquad m \leq 2 n \;.
\]
The logarithmic threshold corrections can be resummed using the direct QCD techniques~\cite{Sterman:1986aj, Catani:1989ne} or in the framework of soft-collinear effective field theories~\cite{Becher_2006}.
In the following we briefly review the threshold resummation formalism applied to the squark-(anti)squark production in direct QCD.

\subsection{General framework}
The resummation of the soft-gluon contributions is performed after taking a Mellin transform (indicated by a tilde) of the hadronic cross section,
\begin{align}
  \label{eq:res_xsec}
  \tilde\sigma_{h_1 h_2 \to \sq\sq^{(*)}}\bigl(N, \{m^2\}\bigr)
  &\equiv \int_0^1 d\rho\;\rho^{N-1}\;
           \sigma_{h_1 h_2\to \sq\sq^{(*)}}\bigl(\rho,\{ m^2\}\bigr) \nonumber\\
  &=      \;\sum_{i,j} \,\tilde f_{i/{h_1}} (N+1,\mu^2)\,
           \tilde f_{j/{h_2}} (N+1, \mu^2) \,
           \tilde{\hat{\sigma}}_{ij \to \sq\sq^{(*)}}\bigl(N,\{m^2\},\mu^2\bigr)\,,
\end{align}
where $\{m^2\}$ denotes all masses entering the calculations and $\rho$ is a hadronic threshold variable, $\rho=4m_\sq^2/S$. Partonic quantities are indicated by a hat symbol. The logarithmic contributions become large when the partonic centre-of-mass energy $\sqrt {\hat{s}}$ approaches $2 m_{\sq}$, i.e.\ in this limit, the partonic threshold variable $\hat{\rho}=4m_{\sq}^2/\hat{s}$ behaves as $\hat\rho \to 1$. The corresponding limit in Mellin space is $N\to\infty$.

The all-order summation of logarithmic terms is based on the factorisation of the hard, (soft)-collinear, and wide-angle soft modes in the cross section~\cite{Sterman:1986aj,Catani:1989ne,Bonciani:1998vc,Contopanagos:1996nh,
  Kidonakis:1998bk,Kidonakis:1998nf, Beneke:2010da}. Near threshold, the resummed partonic cross section up to NNLL accuracy has the form:
\begin{align}
  \label{eq:res_exp}
  \tilde{\hat{\sigma}}^{\mathrm{(res)}} _{ij\rightarrow \sq\sq^{(*)}}\bigl(N,\{m^2\},\mu^2\bigr)
  &=\sum_{I}\,
      \tilde{\hat{\sigma}}^{(0)}_{ij\rightarrow \sq\sq^{(*)},I}\bigl(N,\{m^2\},\mu^2\bigr)\,\left(1+\frac{\als}{\pi}\; C^{\mathrm{(1)}}_{ij\to\sq\sq^{(*)},I}(N,\{m^2\},\mu^2)\right)\;\nonumber\\
  &\qquad\quad \times\,\exp\Big[L g_1(\als L) + g_{2,I}(\als L) +\als g_{3,I}(\alpha_{\mathrm {s}}L) \Big]\,,
\end{align}
where $\tilde{\sigma}^{(0)}_{ij \rightarrow \sq\sq^{(*)},I}$ denotes the colour-decomposed LO cross section in Mellin-moment space. The colour label $I$ corresponds to an irreducible representation of the colour structure of the process. The one-loop matching coefficient $C^{\mathrm{(1)}}_{ij\to\sq\sq^{(*)},I}$ collects all $\mathcal{O}(\als^3)$ non-logarithmic (in $N$) contributions which do not vanish at threshold. The exponent in the second line of Eq.~\eqref{eq:res_exp} captures all dependence on the large logarithm $L=\log(N)$. Reaching the NLL accuracy requires knowledge of the $g_1$ and $g_2$ functions in the exponential. The NNLL accuracy additionally requires knowledge of $g_3$, as well as the one-loop matching coefficient $C^{\mathrm{(1)}}$ which is not included at NLL. While the exponential factor in Eq.~\eqref{eq:res_exp} is the same as in the case of the MSSM and the functions $g_i$ are well known and can be found in e.g.~\cite{Kulesza:2009kq, Beenakker:2011sf}, the calculation of  $C^{\mathrm{(1)}}$ needs to be carried out anew for the MRSSM.

Our calculations include the soft-gluon corrections up to the NNLL accuracy for the colour channels for which it is possible to produce the sparticle pair with zero angular momentum, i.e.\ in the $s$-wave. As seen in Sec.~\ref{sec:MRSSM}, this is the case for the singlet $\mathbf{1}$ and the symmetric octet $\mathbf{8_S}$ colour configurations for the squark-antisquark production in the $gg$ channel. Otherwise, analogously to the approach adopted for the MSSM~\cite{Beenakker:2014sma, Beenakker:2016gmf, Beenakker:2016lwe}, for higher partial wave contributions resummation is performed only up to the NLL accuracy.

The hadronic cross section $\sigma_{h_1 h_2 \to \sq\sq^{(*)}}$ in physical momentum space can be obtained by performing the inverse Mellin transform. To retain full information contained in the complete NLO cross section~\cite{Diessner:2017ske}, we combine the NLO and NNLL results through a matching procedure that avoids double counting of the NLO terms:
\begin{align}
 \label{eq:matching}
  &\sigma^{\mathrm{(NLO+(N)NLL)}}_{h_1 h_2 \to \sq\sq^{(*)}}\bigl(\rho, \{m^2\},\mu^2\bigr)
  = \sum_{i,j}\,\int_\mathrm{CT}\,\frac{dN}{2\pi \ii}\,\rho^{-N} \tilde f_{i/h_1}(N+1,\mu^2)\tilde f_{j/h_{2}}(N+1,\mu^2) \\
&\qquad\times       \left[\tilde{\hat{\sigma}}^{\mathrm{(res)}}_{ij\to \sq\sq^{(*)}}\bigl(N,\{m^2\},\mu^2\bigr)
             \,-\, \left.\tilde{\hat{\sigma}}^{\mathrm{(res)}}_{ij\to \sq\sq^{(*)}}\bigl(N,\{m^2\},\mu^2\bigr)
       \right|_{\scriptscriptstyle{\mathrm{(NLO)}}}\, \right]  +\sigma^{\mathrm{(NLO)}}_{h_1 h_2 \to \sq\sq^{(*)}}\bigl(\rho, \{m^2\},\mu^2\bigr),  \nonumber
\end{align}
where $\mathrm{``res"}$ refers to either NNLL or NLL, depending on the dominant partial wave contribution.
The integral of the inverse Mellin transform is calculated along a contour, denoted here as CT, according to the ``minimal prescription'' of \cite{Catani:1996yz}.

\subsection{Calculation of the hard-matching coefficient}
As discussed above, in order to perform NNLL resummation for the $\sq\sq^*$ production in the MRSSM, we need to calculate the corresponding one-loop matching coefficients $C_{gg \to \sq\sq^*,I}^{\mathrm{(1)}}$ for $I ={\mathbf{1}, \mathbf{8_S}}$. They can be computed either by expanding  the full NLO partonic cross section in the threshold variable $\beta$, or by separately calculating the virtual and real corrections in the threshold limit, 
\begin{equation}
\hat{\sigma}^{(1,\myth)}_I= \hat{\sigma}^{(\text{V}, \myth)}_I+\hat{\sigma}^{(\text{R},\myth)}_I. \nonumber
\end{equation}
Most importantly, the latter approach, discussed in detail in~\cite{Beenakker:2011sf}, does not rely on the knowledge of the analytical expression for the full cross section. The calculation of the $C^{\mathrm{(1)}}$ coefficient requires determining  terms that are proportional to $\beta^0$, $\beta$, $\beta\log\left(\beta\right)$, $\beta\log^2\left(\beta\right)$. Terms with higher powers of $\beta$ are suppressed in Mellin space and therefore do not contribute in the threshold limit.

The expression for the real contributions at threshold, derived for any $2 \to 2$ process with massive coloured particles in the final state, can be found in~\cite{Beenakker:2013mva}. However, the virtual contributions to the MRSSM cross sections need to be separately calculated. To this end, we developed a subroutine to colour-decompose the amplitude squared at the loop level inside the \texttt{FeynArts/FormCalc} framework~\cite{Hahn:2000kx, Hahn:1998yk}. The model file used for the MRSSM was generated by \texttt{SARAH}~\cite{Staub:2009bi, Staub:2010jh, Staub:2012pb, Staub:2013tta} and the one-loop counterterms were included by hand~\cite{Diessner:2017ske}. The colour-decomposed amplitude squared was then expressed in terms of masses, Mandelstam variables and scalar integrals, paying special attention to the conventions used in the expressions for scalar integrals. In the next step, we replaced the Passarino-Veltman integrals with their analytical expansions using \texttt{FeynCalc, FeynHelpers, PackageX}~\cite{Mertig:1990an, Shtabovenko:2016sxi, Shtabovenko:2020gxv, Shtabovenko:2016whf, Patel:2016fam, Ellis:2007qk}, and integrated over the phase space. Calculated in the conventional dimensional regularization scheme, the factors that multiply the scalar integrals have no negative powers of $\beta$, i.e.\ their expansions start at ${\mathcal O} (\beta^0)$.
Since the phase-space integration brings terms of $\mathcal{O}(\beta)$ and higher, one does not need to expand the scalar integrals beyond $\mathcal{O}\left(\beta^0\right)$ to find out the contributions to the cross section up to ${\mathcal O} (\beta)$.
Special attention has to be paid to the Coulomb integrals which are proportional to inverse powers of $\beta$ and cancel the $\propto\beta$ suppression from the phase-space integration. Consequently, when transforming to Mellin-moment space, they are $N$-dependent, and we thus subtract the Coulomb integrals for the purpose of calculating only hard-matching coefficients, i.e.\ the $N$-independent contributions to the matching coefficients. In the result of Eq.~\eqref{eq:res_exp}, the Coulomb contributions are again added.
After combining the real and the virtual threshold corrections, the result must be free of divergences, which provides a strong check on the final result. The one-loop matching coefficients are now obtained by dividing the Mellin-transformed result by the LO threshold cross section in Mellin space and considering only the terms that do not contain powers of $\log N$. One can schematically represent this by:
\begin{equation}
	\frac{\als}{\pi} C^{\mathrm{(1)}}_{gg\to\sq\sq^{*},I}= \left.\dfrac{\tilde{\hat{\sigma}}^{(1,\myth)}_{gg\to\sq\sq^{*},I}}{\tilde{\hat{\sigma}}^{(0,\myth)}_{gg\to\sq\sq^{*},I}}\right|_{\log N-\text{independent}}. \nonumber
\end{equation}
We have checked that this approach reproduces the known analytical expressions for $C_{ij \to \sq\sq^{*},I}^{(1)}$ coefficients in the MSSM~\cite{Beenakker:2013mva}.

\begin{figure}[tp]
  \centering
 \scalebox{0.85}{
  \begin{subfigure}{0.475\textwidth}
    \includegraphics[width=\linewidth]{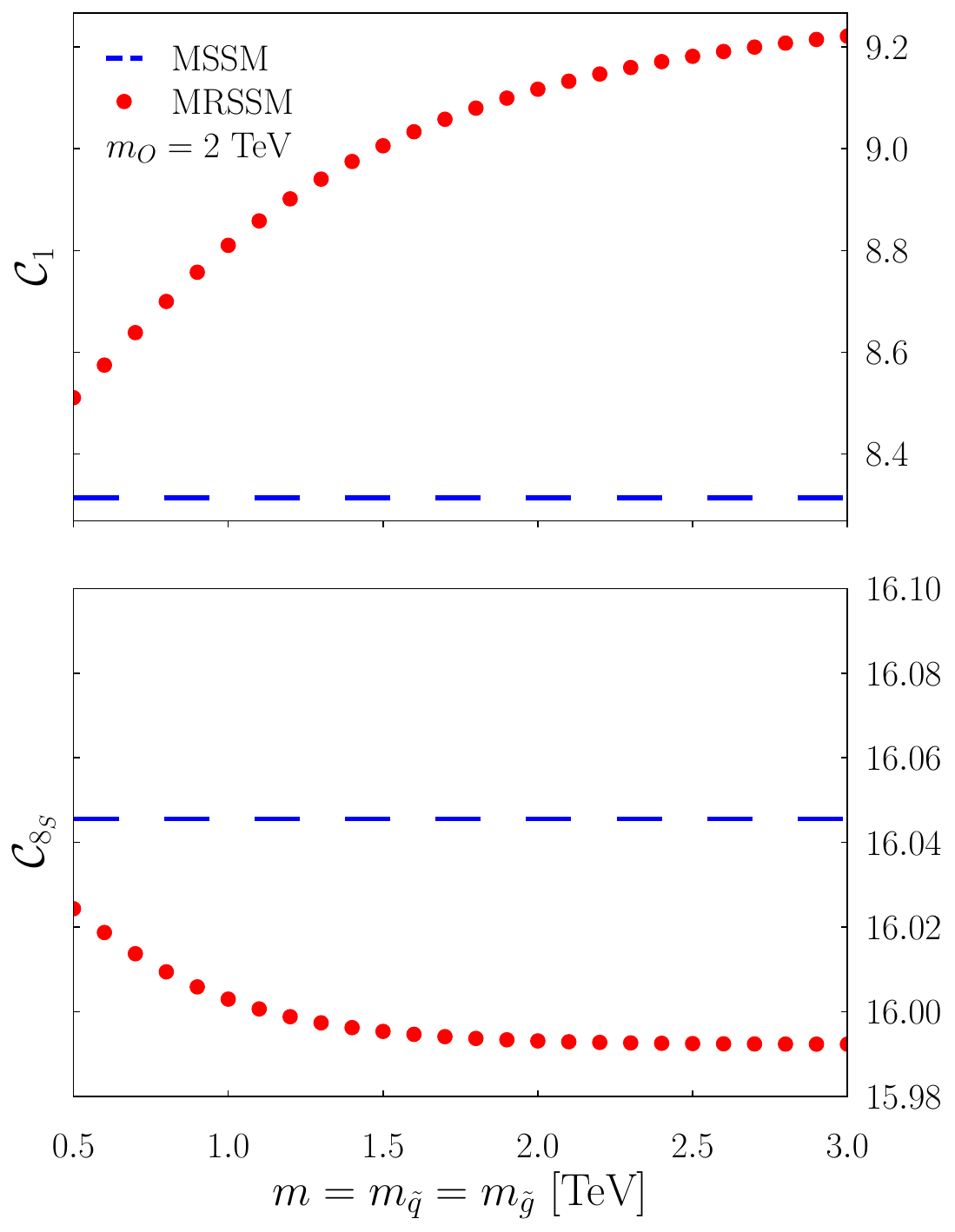}
    \label{fig: C coefficient mO 2}
  \end{subfigure}
  \hfill
  \begin{subfigure}{0.475\textwidth}
    \includegraphics[width=\linewidth]{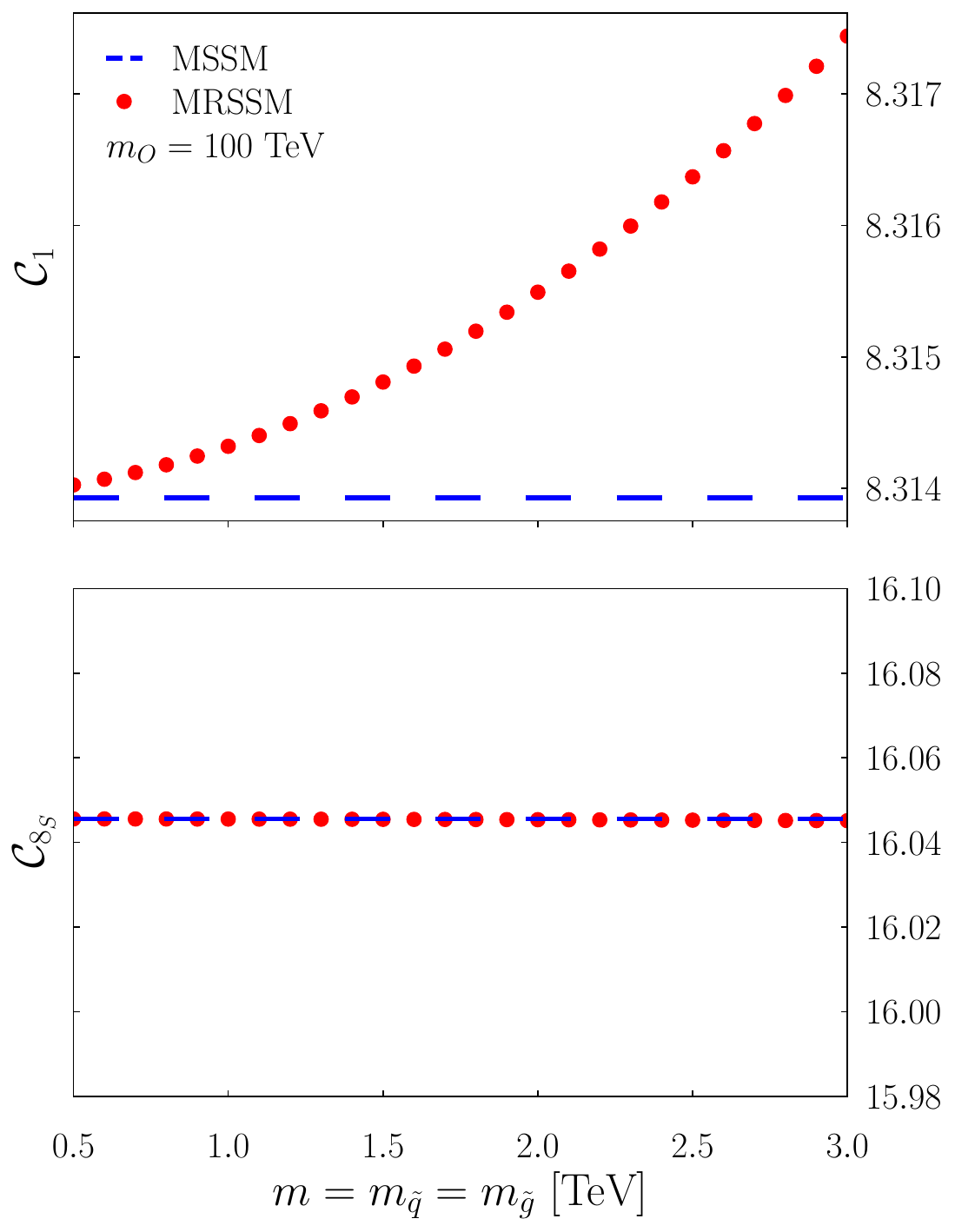}
    \label{fig: C coefficient mO 100}
  \end{subfigure}}
  \caption{Mass dependence of the hard-matching coefficients $\mathcal{C}_{gg \to \sq\sq^*,\mathbf{1}}^{(1)}$ (top) and $\mathcal{C}_{gg \to \sq\sq^*,\mathbf{8_S}}^{(1)}$ (bottom) in the MRSSM and the MSSM for two sgluon masses $m_O=2$ TeV (left) and  $m_O=100$ TeV (right), calculated setting $\mu_F=\mu_R=m$.}
  \label{fig: C coefficient}
\end{figure}

The complete analytical expressions for the one-loop matching coefficients $C_{gg \to \sq\sq^*,I}^{(1)}, \ I=\mathbf{ 1, 8_S}$ including the split into Coulomb corrections and the hard-matching coefficients $\mathcal{C}_{gg \to \sq\sq^*,I}^{(1)}$ can be found in App.~\ref{appendix: ccoeffs}. The behaviour of the latter as a function of $m=\msq$ for two values of the sgluon mass $m_O =2$ TeV and $m_O=100$ TeV, and under the assumption of equal squark and gluino masses $m=\msq=\mgl$, is shown in Fig.~\ref{fig: C coefficient}. We also show the corresponding hard-matching coefficients in the MSSM, which reduce to a constant for $\msq=\mgl$ and $\mu_F=\mu_R=m$. The biggest relative difference between the one-loop hard-matching coefficients in the MRSSM and the MSSM can be observed for the colour singlet production when the sgluon masses are small. As $m$ decreases, the differences between results in the two models become smaller, since in the limit $m=\mgl \to 0$ the nature of the gluino does not play a role and the cross sections in the two models are equal. This is the same effect as already observed for the LO cross sections in Sec.~\ref{sec:MRSSM}, see Fig.~\ref{fig: LOresultsat136}. The MSSM results are also recovered in the limit of large sgluon masses, where the sgluon degree of freedom decouples. For the range of masses considered in Fig.~\ref{fig: C coefficient}, this effect is particularly visible in the $\mathbf{8_S}$ channel.

\section{Numerical results}\label{sec:results}
In the following, we present numerical predictions for $\sq\sq^*$ and $\sq\sq$ production\footnote{We always imply the Hermitian conjugated process, so when discussing $\sq\sq$ production, we actually consider the sum of $\sq\sq$ and $\sq^*\sq^*$.} in proton-proton collisions at the LHC with a centre-of-mass energy of $\sqrt S=13.6$ TeV, obtained using the \texttt{PDF4LHC21\_40\_pdfas} Hessian PDF set \cite{PDF4LHCWorkingGroup:2022cjn} accessed via the \texttt{LHAPDF} library \cite{Buckley:2014ana}. Unless otherwise noted, the renormalisation and factorisation scales have been chosen equal, $\mu \equiv \mu_R = \mu_F$, and set to the central scale choice $\mu = \mu_0 = m_\sq$, and a top-quark mass of $m_t=172$ GeV is used. We consider the left- and right-handed superpartners of the light quarks $u$, $d$, $c$, $s$, $b$ to be degenerate in mass\footnote{For non-degenerate squark masses, we expect a similar behaviour as for the MSSM, studied e.g.\ in \cite{Goncalves-Netto:2012nvl,Gavin:2013kga,Gavin:2014yga}. There, it was found that the difference between NLO $K$-factors for degenerate and non-degenerate squark mass spectra is negligible. Thus, to a good approximation, higher-order results in the non-degenerate case can be obtained by rescaling the LO cross section with non-degenerate masses by the NLO and (N)NLL $K$-factors of the degenerate case.}, and the results are thus given assuming a 10-fold squark degeneracy with $\sq = \tilde{u}_L$, $\tilde{u}_R$, $\tilde{d}_L$, $\tilde{d}_R$, $\tilde{c}_L$, $\tilde{c}_R$, $\tilde{s}_L$, $\tilde{s}_R$, $\tilde{b}_L$, $\tilde{b}_R$, excluding the production of stops as the partners of top quarks, corresponding to the treatment in the MSSM calculations\footnote{Since we are not considering top quarks in the initial state, at LO, $t$-channel diagrams with gluino exchange are absent for stop production. Thus, we do not consider them with the other squark flavours. The dependence on the gluino and sgluon masses is only introduced at higher orders, therefore we expect their impact to be small and the process to behave similarly to the MSSM \cite{Beenakker:2016gmf}.} \cite{Beenakker:2014sma, Beenakker:2016gmf, Beenakker:2016lwe}. The NLO cross sections are calculated using \texttt{MadGraph5\_aMC@NLO} v2.9.16 \cite{Alwall:2014hca}, with virtual matrix elements provided by \texttt{GoSam} \cite{GoSam:2014iqq} based on the MRSSM \texttt{UFO} \cite{Degrande:2011ua,Staub:2012pb} model file. The required renormalisation constants are added by hand to the Fortran code generated by \texttt{GoSam} (for a comprehensive explanation of this procedure consult Ref.~\cite{Diessner:2017ske}). The on-shell gluino singularity in real emission contributions is handled using diagram subtraction via \texttt{MadSTR} \cite{Frixione:2019fxg} (variant \texttt{istr=3} with \texttt{str\_include\_pdf=True} and \texttt{str\_include\_flux=True}). This semi-automated calculation was cross-checked against our public code \texttt{RSymSQC}, which calculates cross sections for respective processes based on an analytic calculation \cite{RSymSQCD}\footnote{It is planned that a manual for it will be published shortly after the present work.}. The corrections due to threshold resummation have been calculated and cross-checked using two independent in-house codes.

\begin{figure}[tp]
 \begin{subfigure}{\textwidth}
 \centering
 \includegraphics[width=.5\textwidth]{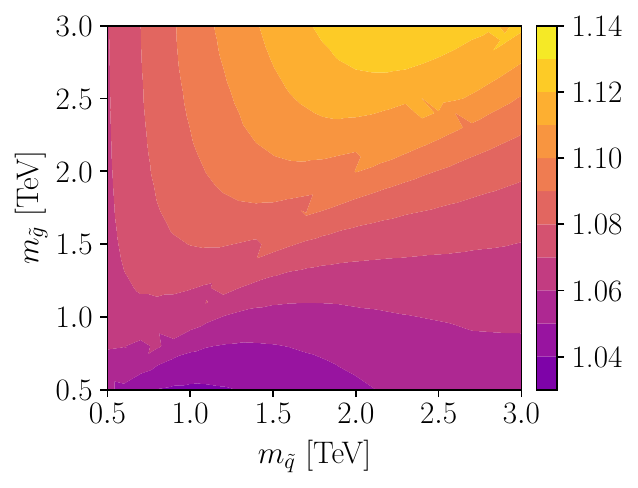}\includegraphics[width=.5\textwidth]{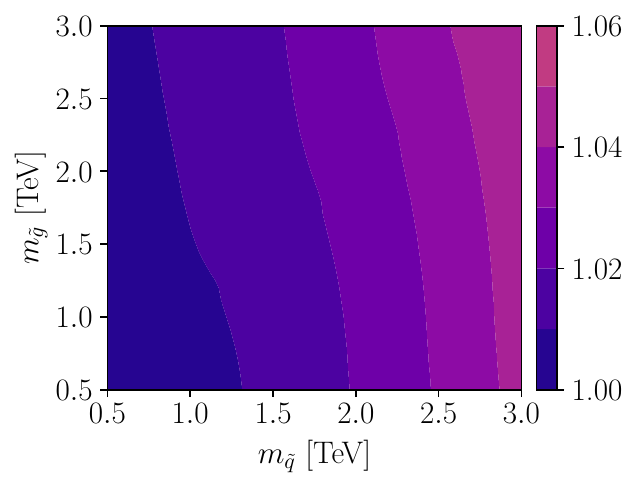}
 \end{subfigure}
 \begin{subfigure}{\textwidth}
 \centering
 \includegraphics[width=.5\textwidth]{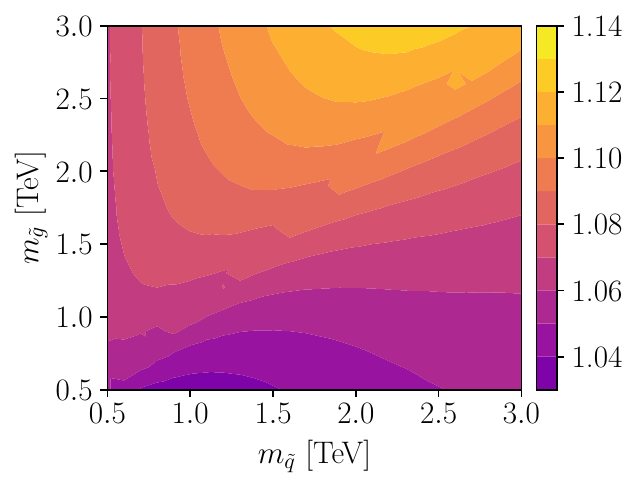}\includegraphics[width=.5\textwidth]{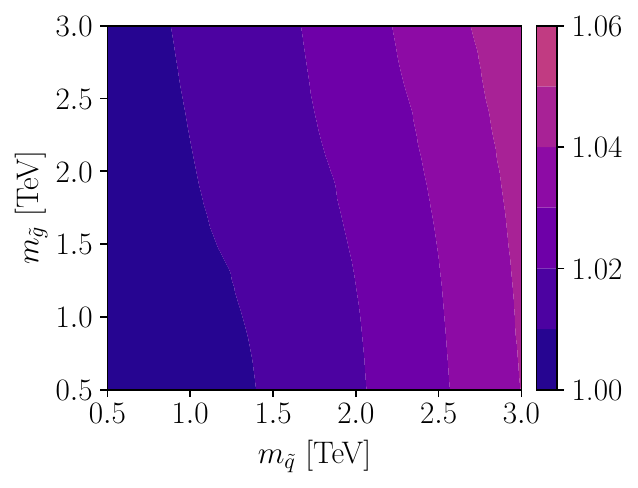}
 \end{subfigure}
 \begin{subfigure}{\textwidth}
 \centering
 \includegraphics[width=.5\textwidth]{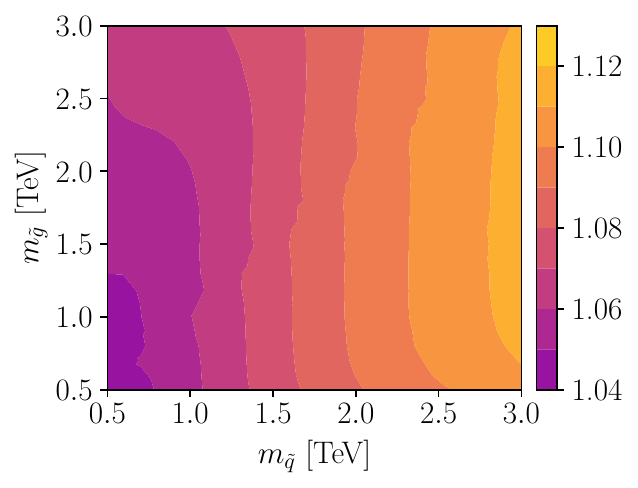}\includegraphics[width=.5\textwidth]{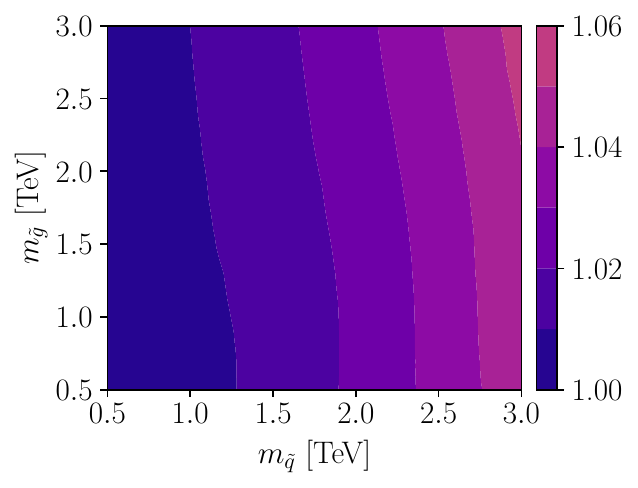}
 \end{subfigure}
 \caption{$K_{\mathrm {NNLL}}$ factors for $\sq\sq^{*}$ production (left column) and $K_{\mathrm {NLL}}$ factors for $\sq\sq$ production (right column), as defined in Eq.~\eqref{eq:kfactor}. Shown are the MRSSM results evaluated for a sgluon mass of $m_O=2$~TeV (top row), for $m_O=100$~TeV (middle row), and the MSSM results (bottom row).}
 \label{fig:K-NNLL}
\end{figure}
We begin by analysing the impact of soft-gluon corrections on the theoretical predictions for squark production. In Fig.~\ref{fig:K-NNLL} we show the (N)NLL $K$-factors, defined as the ratio of the resummed and matched cross section according to Eq.~\eqref{eq:matching} over the NLO cross section,
\begin{equation}\label{eq:kfactor}
	K_{\mathrm{(N)NLL}} \equiv \frac{\sigma^{\mathrm {(NLO+(N)NLL)}}}{\sigma^{\mathrm{ (NLO)}}}\,,
\end{equation}
for $\sq \sq^*$ and $\sq\sq$ production in the MRSSM and MSSM, as a function of $\msq$ and $\mgl$. We study each process in the MRSSM for two values of the sgluon mass, $m_O = \{2, 100\}$~TeV, probing two opposite ends of the sgluon mass spectrum. As explained above, the accuracy of resummation is different for the two processes due to the difference in the composition of the cross section in terms of partial waves. It reaches NLO+NNLL for the $\sq \sq^*$ process, while the $\sq\sq$ production is considered at NLO+NLL. As expected, we see that soft-gluon corrections get larger with increasing squark masses. The NNLL corrections can increase the $\sq\sq^*$ cross section by up to 14\%, both in the MRSSM and MSSM. However, the MRSSM and the MSSM $K$-factors for the $\sq\sq^*$ process differ substantially in their dependence on the squark and gluino mass. While the $\sq\sq^*$ MSSM $K$-factor almost does not depend on $\mgl$, the MRSSM $K$-factor appears to have a stronger dependence on $\mgl$ than on $\msq$. This means that at NNLL, an MRSSM correction of comparable size to the one in the MSSM can only be obtained if the gluino mass is sufficiently high. For $\sq\sq$ production in the MRSSM, with the dominant contribution to the LO cross section in the $p$-wave channel and consequently, the resummed correction being only of NLL order, the change in the cross section amounts to at most a few per cent for $\msq$ and $\mgl$ up to 3 TeV. The relative soft-gluon corrections to the $\sq\sq$ and $\sq\sq^*$ MRSSM cross sections do not significantly depend on the sgluon mass, in agreement with the sgluon mass only entering in the genuine one-loop diagrams.

\begin{figure}[tp]
  \begin{subfigure}{0.50\textwidth}
 \includegraphics[width=1.\textwidth]{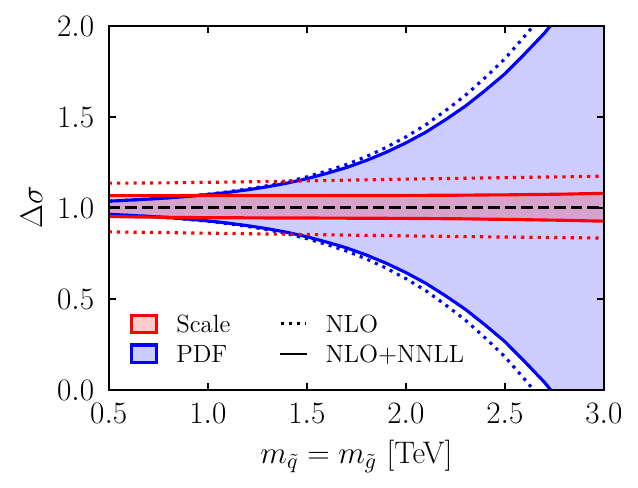}
 \label{fig: errors for sq sqb}
 \end{subfigure}%
 \begin{subfigure}{0.50\textwidth}
 \includegraphics[width=1.\textwidth]{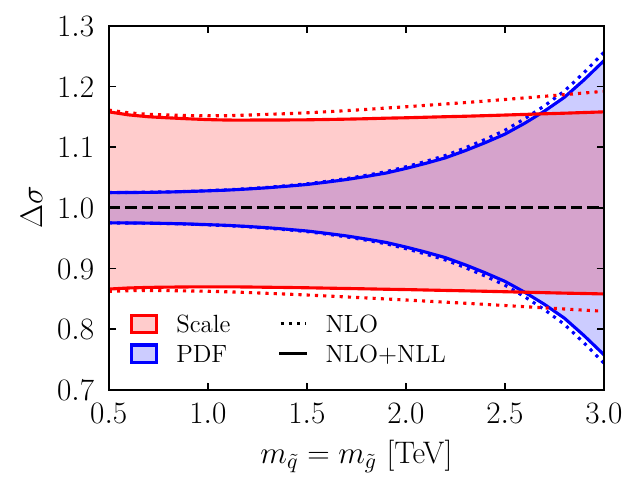}
 \label{fig: errors for sq sq}
 \end{subfigure}
 \caption{Scale and PDF+$\als$ uncertainties for $\sq\sq^*$ production at NLO+NNLL (left) and $\sq\sq$ production at NLO+NLL (right) in the MRSSM, with the sgluon mass being fixed to $m_O=100$~TeV. $\Delta\sigma$ is defined as the relative uncertainty with respect to the central cross-section prediction. The scale uncertainty shown here is obtained via the 3-point method, see the text below.}
 \label{fig:totaluncertainty}
\end{figure}
Apart from providing a positive correction to the total cross section, resummation impacts the theoretical prediction by reducing the size of the theoretical error due to scale variation. We illustrate this effect for the MRSSM cross sections in Fig.~\ref{fig:totaluncertainty}, where we only show the case of $m_O=100$ TeV as the uncertainties for $m_O=2$ TeV look almost identical, with an overall slightly reduced size. The scale error is calculated using the 3-point method, i.e., considering the variation of $\mu$ around the central scale $\mu_0=\msq$ between $0.5 \mu_0$ and $2 \mu_0$. This method of estimating the scale error corresponds to the scale error estimate in the MSSM provided by the NNLL-fast code~\cite{Beenakker:2016lwe}. Additionally, in Fig.~\ref{fig:totaluncertainty}, we also show the size of the theoretical error due to the PDF uncertainties including also a variation of the value of $\als$. Compared to the NLO cross section, the scale uncertainty of the $\sq\sq^*$ cross section at NLO+NNLL is reduced by almost a half for all values of $\msq=\mgl$. At small mass values, the scale error is now of a similar size as the PDF error, while at NLO it constitutes the dominant source of the theory uncertainty. In line with the resummation corrections being significantly smaller at NLL and for $p$-wave channels, the reduction of the scale error is much less pronounced for the $\sq\sq$ process. At high masses, the PDF error begins to dominate, although, due to the presence of the $q\bar q$ channel at LO and thus the appearance of sea-quark PDFs, it happens at lower mass values for the $\sq\sq^*$ rates than for $\sq\sq$, which only receives a contribution from the $qq$ initial state, dominated by valence-quark PDFs, at LO. As both the fixed order and resummed cross sections are calculated with the same NNLO PDF4LHC21 set, the difference in the size of the PDF error for the two results is understandably modest. We note in passing that in case of unequal squark and gluino masses, the PDF error mainly depends on the squark masses. Therefore, its dependence on the gluino mass is moderate.
In fact, if we fix $\mgl=3$ TeV, and therefore just consider the dependence on the squark mass, the PDF error is almost exactly the same as in the equal-mass case.

\begin{figure}[tp]
 \begin{subfigure}{\textwidth}
 \centering
 \includegraphics[width=\textwidth]{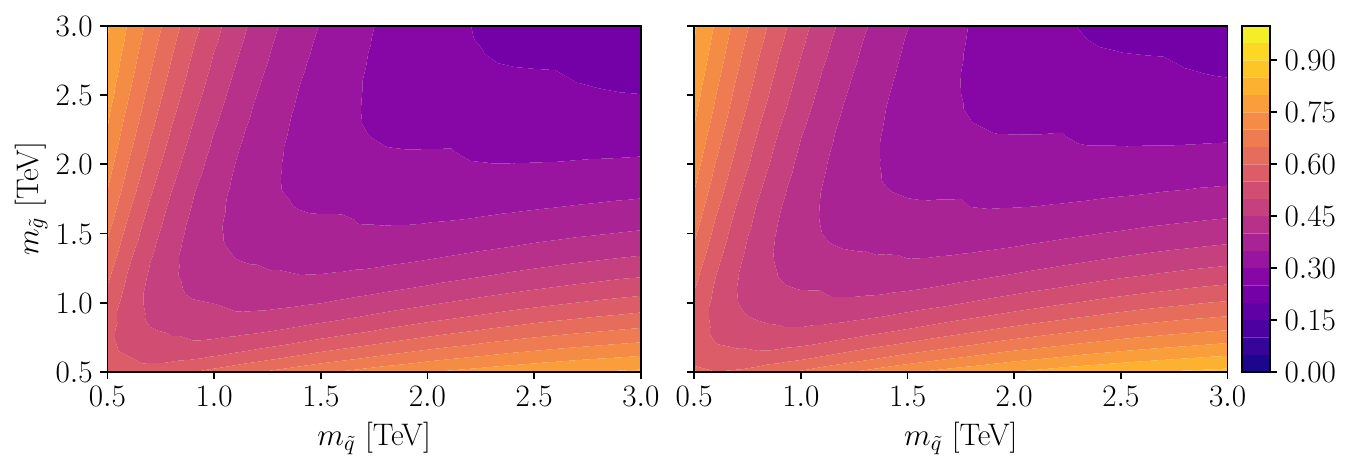}
 \label{fig:ratiosqsqb}
 \end{subfigure}
 \begin{subfigure}{\textwidth}
 \centering
 \includegraphics[width=\textwidth]{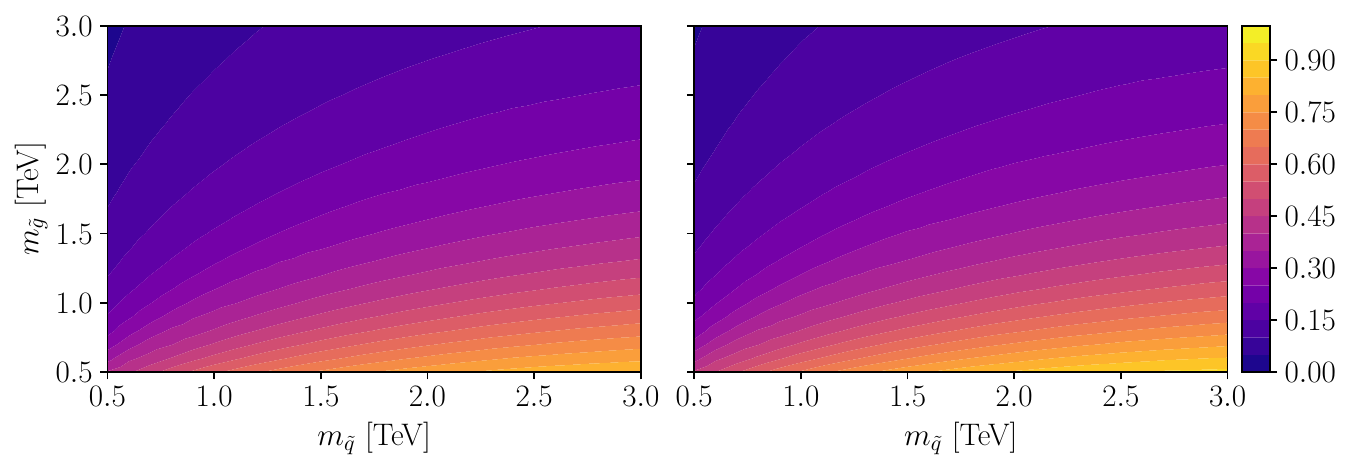}
 \label{fig:ratiosqsq}
 \end{subfigure}
 \caption{
 Ratios of MRSSM cross sections over the MSSM ones, computed at the same accuracy. For $\sq\sq^*$ production (top row), the cross sections are calculated at NLO+NNLL, while for $\sq\sq$ production (bottom row), the calculation is performed at NLO+NLL. The ratios are shown for two sgluon mass configurations for the MRSSM predictions: $m_O=2$~TeV (left column), and $m_O=100$~TeV (right column).
 }
 \label{fig:ratioxsec}
\end{figure}
We then move onto discussing the differences between the resummation-improved predictions for squark pair production in both models. Fig.~\ref{fig:ratioxsec} shows the ratio of the NLO+NNLL cross sections in the MRSSM and MSSM for $\sq\sq^*$ production (top row), and the ratio of the NLO+NLL cross sections in the MRSSM and MSSM for $\sq\sq$ production (bottom row), respectively. For both processes, the MRSSM cross section is drastically reduced compared to the MSSM. The stronger effect of the two is present for the $\sq\sq$ process, where for $\mgl \sim 3$ TeV the cross section is reduced by almost two orders of magnitude. While the MRSSM $\sq\sq$ cross section departs most rapidly from the MSSM one with increasing $\mgl$, for $\sq\sq^*$ this happens when both $\msq$ and $\mgl$ grow bigger simultaneously. In the latter case, the NLO+NNLL cross section in the MRSSM is about five times smaller than in the MSSM for $\msq$ and $\mgl$ around 3 TeV. The origin of this behaviour can be traced back to LO, discussed in Sec.~\ref{sec:MRSSM}. In particular, the extreme suppression of the $\sq\sq$ cross section mentioned there can also be observed in Fig.~\ref{fig:ratioxsec}. It is driven by the $t$-channel gluino exchange diagram shown in Fig.~\ref{figwithtwotreelevelandoneloopdiagramsinvolvingsgluon} which yields vastly different results in the two models, resulting at large $\mgl$ in a $m_{\tilde g}^{-4}$ suppression in the MRSSM, as opposed to the $m_{\tilde g}^{-2}$ suppression in the MSSM. For $\sq\sq^*$, on the other hand, the diagram with $t$-channel gluino exchange is suppressed for infinite gluino masses, so that the cross sections become equal in the two models, as shown in the right-most plot of Fig.~\ref{fig: LOresultsat136} for LO results. We note that the higher-order corrections to the $\sq\sq^*$ production cross sections exhibit the same behaviour.

\begin{figure}[tp]
 \centering
 \includegraphics[width=.77\textwidth]{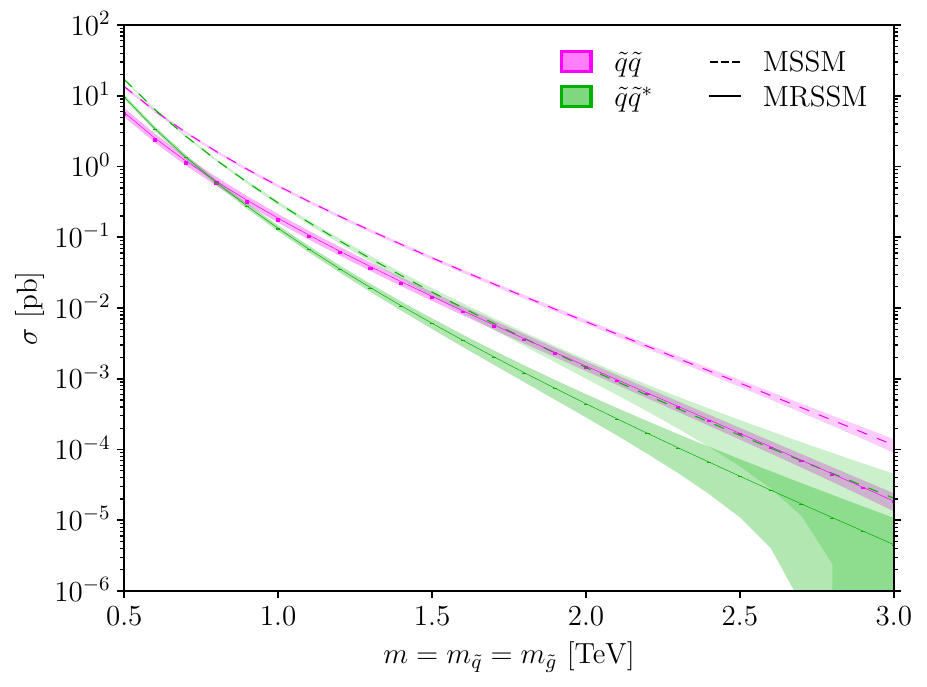}
 \caption{Total cross-section predictions for $\sq\sq^*$ and $\sq\sq$ production at the LHC with \mbox{$\sqrt{S} = 13.6$} TeV confronting the two models MSSM and MRSSM at their best accuracy as explained in the text. The error bands denote the total theoretical uncertainty due to scale variation with the 3-point method and the PDF+$\als$ error, added in quadrature. The MRSSM predictions have been computed with a sgluon mass of $m_O = 100$~TeV, and the variation due to changing the sgluon mass to $m_O = 2$~TeV is shown as one-sided error bars below the central values.}
 \label{fig: cross sections}
\end{figure}
Finally, in Fig.~\ref{fig: cross sections}, we present the total cross-section predictions in the MRSSM for $\sq\sq^*$ and $\sq\sq$ production at their best available accuracies NLO+NNLL and NLO+NLL, respectively, and compare to the state-of-the-art results in the MSSM from NNLL-fast \cite{Beenakker:2016lwe}, evaluated at NNLO$_\text{Approx}$+NNLL accuracy including the resummation of Coulomb-gluon exchange and bound-state contributions, and computed with PDF4LHC21 PDFs and a centre-of-mass energy of $\sqrt{S} = 13.6$~TeV. For this plot, the squark and gluino masses have been set equal, $m \equiv m_\sq = m_\gl$. We show the central cross sections for the LHC with $\sqrt{S} = 13.6$~TeV as a function of the common mass $m$ including total uncertainty bands, consisting of the scale and PDF uncertainties added in quadrature. For the central predictions in the MRSSM, a sgluon mass of $m_O = 100$~TeV has been used. As already discussed, we see that MRSSM cross sections are significantly reduced compared to the case of the MSSM. The higher the common mass $m$, the stronger the reduction in the MRSSM becomes. As shown by Fig.~\ref{fig:ratioxsec}, the reduction of the MRSSM cross sections also depends on the squark-gluino mass splitting. For the MRSSM results, the effect of the variation of the sgluon mass from $m_O = 100$~TeV to $m_O = 2$~TeV is shown in Fig.~\ref{fig: cross sections} as one-sided error bars below the central values\footnote{As cross sections with $m_O = 100$~TeV are typically larger than with $m_O = 2$~TeV, the error bars are plotted only in the direction of smaller cross sections.}. While the effect is generally rather small, in particular for large squark and gluino masses, the variation can be of the size of the combined scale and PDF uncertainty for small squark and gluino masses. The difference between the two sgluon masses can impact the $\sq\sq$ cross sections between 5\% and 17\%, while the effect is smaller for $\sq\sq^*$ and reaches only from 0\% up to 8\%. We note that the dependence on the sgluon mass is mainly driven by the behaviour of the NLO cross section, as beyond NLO, the sgluon mass only appears in the hard-matching coefficients for $\sq\sq^*$ of the non-dominant $gg$ initial-state channels.

The results of Figs.~\ref{fig:ratioxsec} and \ref{fig: cross sections} highlight the importance for experimental searches, as a reduced cross section will lead to less stringent mass limits for squarks and gluinos, and subsequently a larger allowed parameter space in the MRSSM than in the MSSM. The size of the uncertainty band in the MSSM is smaller compared to the MRSSM, due to the additional contributions of higher orders taken into account in the resummation procedure in the former case.

\section{Conclusions}\label{sec:conclusions}
The Minimal R-symmetric Supersymmetric Standard Model is an appealing alternative minimal realization of supersymmetry. Not only does it predict a rich electroweak sector, but it also involves a strongly altered coloured sector compared to the usual MSSM, with significant modifications to LHC phenomenology. Most notably, half of the production channels for squarks or antisquarks are forbidden, and the remaining ones are suppressed by the Dirac nature of the gluino mass.

The present paper presents state-of-the-art computations for the LHC production of squarks in the $\sq\sq$ and $\sq\sq^*$ channels. It takes into account fixed order NLO corrections and threshold resummation either at the NLL or NNLL accuracy. 
While the NLO corrections have been already available~\cite{Diessner:2017ske}, the calculation of the resummation-improved cross sections presented here is new. Apart from increasing the predictions for the total cross sections, the threshold corrections reduce the theoretical error due to scale variation, particularly for the $\sq\sq^*$ production.
With this computation, the MRSSM predictions are available at
almost the same accuracy as MSSM predictions implemented in the public
code NNLL-fast~\cite{Beenakker:2016lwe}.

We have presented a detailed comparison between the MRSSM and the MSSM
predictions for LHC squark/antisquark production. The leading order as
well the NLO and also the resummation corrections show characteristic
differences between the two models. Most notably, the MRSSM cross
sections are lower by up to two orders of magnitude in parameter
regions of interest. In addition, the suppression depends strongly on
the overall mass scale as well as on the mass splitting between
squarks and gluinos.

Because of these differences, clearly lower squark masses are compatible with LHC searches in the MRSSM compared to the MSSM. Preliminary analyses using recastings of LHC data have shown that around 500--600 GeV lighter squarks are possible in the MRSSM~\cite{Kribs:2013oda,Diessner:2019bwv}.

It would be valuable to delineate the allowed region of parameter space in the MRSSM more accurately, to determine the additional viable squark mass range in presence of R-symmetry. The computations of the present paper enable such a dedicated analysis of actual LHC data for squark searches in the MSSM and the MRSSM in parallel, using similarly precise theory predictions including NLO and (N)NLL corrections.

The cross sections for squark-antisquark and squark-squark production in the MRSSM including soft-gluon corrections up to (N)NLL as calculated in this paper are publicly available, for centre-of-mass energies of $\sqrt{S} = 13$~TeV and $\sqrt{S} = 13.6$~TeV, as a numerical code package from the NNLL-fast website:
\begin{center}
	\url{https://www.uni-muenster.de/Physik.TP/~akule_01/nnllfast}
\end{center}

\section*{Acknowledgements}
WK is supported by the National Science Centre (Poland) grant 2020/\allowbreak38/\allowbreak E/\allowbreak ST2/\allowbreak00126 and by the German Research Foundation (DFG) under grant numbers STO 876/4 and STO 876/2.
FF acknowledges support from the DFG Research Training Group ``GRK
2149: Strong and Weak Interactions - from Hadrons to Dark Matter'' and by the German Research Foundation (DFG) under grant number STO 876/4.

The authors are grateful to the Centre for Information Services and High Performance
Computing [Zentrum f\"ur Informationsdienste und Hochleistungsrechnen (ZIH)] TU Dresden and to the PALMA HPC cluster of the University of M\"unster for providing its facilities for high throughput calculations.

\appendix
\section{Leading order colour decomposed}
\label{appendix: LO colour split}
In this appendix, we list all the used $s$-channel base tensors $c_I$ and
their quadratic Casimir invariant $C_2(R_I)$ for each
colour representation $R_I$ in a general $\mathrm{SU(\SUNN)}$ theory
together with the colour-split LO cross section from Ref.~\cite{Diessner:2017ske}. The base tensors are written in terms of Kronecker deltas in colour space $\delta_{ab}$, the generators of the fundamental representation $T^{c}_{ab}$, the structure constants $f^{abc}$ and their symmetric counterparts $d^{abc}$.
The colour labels $a_1$ and $a_2$ denote the initial-state particles while $a_3$ and $a_4$ refer to the final-state particles~\cite{Beenakker:2013mva}. 
To keep the expressions short, we use the following definitions:
\begin{align}
&\beta = \sqrt{1-\frac{4 \msqq}{\hat{s}}}, &&L_1= \ln\left( \frac{\hat{s}+2m_-^2 - \hat{s}\beta}{\hat{s}+2m_-^2 + \hat{s}\beta}\right);\nonumber\\
&m_-^2=\mglq - \msqq,  &&m_+^2= \mglq + \msqq.\nonumber
\end{align}
The squark-antisquark production in the quark channel can be decomposed as:
\begin{equation*}
q_i(a_1)\overline{q}_j(a_2)\rightarrow \sq(a_3)\sq^{*}(a_4)\qquad \mathbf{\SUNN} \otimes \mathbf{\overline{N}_c}= \mathbf{1}\oplus \left(\mathbf{\SUNN^2}-\mathbf{1}\right); \\
\end{equation*}
\begin{align*}
c_{1} &=\dfrac{1}{\SUNN}\delta_{a_1a_2}\delta_{a_3a_4}; &\qquad C_2\left(R_{1}\right)&= 0;\\
c_{2} &=2T^{c}_{a_2a_1}T^{c}_{a_3a_4}; &\qquad C_2\left(R_{2}\right)&=\SUNN.
\end{align*}
In $\mathrm{SU(3)}$, $R_1=\mathbf{1}$ and $R_2=\mathbf{8}$. Using these bases, we can project the LO cross section [Eq.~(3.3) of~\cite{Diessner:2017ske}] onto the two colour representations:
\begin{align}
\hat{\sigma}^{(0)}_{1}&=\dfrac{\pi\als^2}{\hat{s}}\left({\dfrac{(\SUNN^2-1)^2}{2\SUNN^4}}\right)\left[-2\beta-\left(1+\dfrac{2m^2_-}{\hat{s}}\right)L_1\right]; \\
\hat{\sigma}^{(0)}_{2}&=\delta_{ij}\dfrac{\left(n_f-1\right)\pi\als^2}{\hat{s}}\left({\dfrac{\SUNN^2-1}{6\SUNN^2}}\right)\beta^3\notag\\
&\quad {}+\delta_{ij}\dfrac{\pi \als \alssusy}{\hat{s}}\left({\dfrac{\SUNN^2-1}{2\SUNN^3}}\right)\left[\beta\left(1+\dfrac{2m_{-}^2}{\hat{s}}\right)+\left(\dfrac{2m^2_{\gl}}{\hat{s}}+\dfrac{2m^4_-}{\hat{s}^2}\right)L_1\right]\notag\\
&\quad {}+\dfrac{\pi\alssusy^2}{\hat{s}}\left({\dfrac{\SUNN^2-1}{2\SUNN^4}}\right)\left[-2\beta-\left(1+\dfrac{2m^2_-}{\hat{s}}\right)L_1\right].
\end{align}
If one expands near threshold, $\beta\rightarrow0$, one obtains:
\begin{align}
\hat{\sigma}^{(0,\myth)}_{1}&=\pi\als^2\left({\dfrac{(\SUNN^2-1)^2}{2\SUNN^4}}\right)\left[\beta^3\dfrac{2\msqq}{3\left(\mglq+\msqq\right)^2}\right]; \\
\hat{\sigma}^{(0,\myth)}_{2}&=\delta_{ij}\dfrac{\left(n_f-1\right)\pi\als^2}{4\msqq}\left({\dfrac{\SUNN^2-1}{6\SUNN^2}}\right)\beta^3\notag\\
&\quad {}+\delta_{ij}\pi \als \alssusy\left({\dfrac{\SUNN^2-1}{2\SUNN^3}}\right)\left[\dfrac{\beta^3}{3\left(\mglq+\msqq\right)}\right]\notag\\
&\quad {}+\pi\alssusy^2\left({\dfrac{\SUNN^2-1}{2\SUNN^4}}\right)\left[\beta^3\dfrac{2\msqq}{3\left(\mglq+\msqq\right)^2}\right].
\end{align}
Both contributions (corresponding to singlet and octet cross sections in the case $\SUNN=3$) go to zero as $\beta^3$, i.e.\ they behave as a $p$-wave contribution. This constitutes a big difference to the MSSM as the R-symmetry forbids the production of different ``chiralities'', resulting in the suppression at threshold.
The squark-antisquark production in the gluon channel can be decomposed as:
\begin{equation*}
g(a_1)g(a_2)\rightarrow \sq(a_3)\sq^{*}(a_4)\qquad \left(\mathbf{\SUNN^2}-\mathbf{1}\right) \otimes \left(\mathbf{\SUNN^2}-\mathbf{1}\right) = \mathbf{1}\oplus \left(\mathbf{\SUNN^2}-\mathbf{1}\right)_{\text{S}} \oplus \left(\mathbf{\SUNN^2}-\mathbf{1}\right)_{\text{A}}
\end{equation*}
\begin{align*}
c_{1} &=\dfrac{1}{\sqrt{\SUNN\left(\SUNN^2-1\right)}}\delta^{a_1a_2}\delta_{a_3a_4}; &\qquad C_2\left(R_1\right)&= 0;\\
c_{2} &=\sqrt{\dfrac{2\SUNN}{\SUNN^2-4}}d^{a_1a_2c}T^{c}_{a_3a_4}; &\qquad C_2\left(R_{2}\right)&=\SUNN;\\
c_{3} &=\ii\sqrt{\dfrac{2}{\SUNN}} f^{a_1a_2c}T^{c}_{a_3a_4}; &\qquad C_2\left(R_{3}\right)&=\SUNN.
\end{align*}
If one considers $\mathrm{SU(3)}$ then $R_1=\mathbf{1}$, $R_2=\mathbf{8_S}$, $R_3=\mathbf{8_A}$. We can use these bases to project the LO cross section~[Eq. (3.4) of~\cite{Diessner:2017ske}] onto the three colour representations:
\begin{align}
\hat{\sigma}^{(0)}_{1}&=\dfrac{\pi\als^2}{\hat{s}}\left({\dfrac{n_f-1}{\SUNN\left(\SUNN^2-1\right)}}\right)\left[\left(1+\dfrac{4\msqq}{\hat{s}}\right)\beta+\left(\dfrac{4\msqq}{\hat{s}}-\dfrac{8\msqqu}{\hat{s}^2}\right)\ln\left(\dfrac{1-\beta}{1+\beta}\right)\right]; \\
\hat{\sigma}^{(0)}_{2}&=\dfrac{\pi\als^2}{\hat{s}}\left({\dfrac{\left(n_f-1\right)\left(\SUNN^2-4\right)}{2\SUNN\left(\SUNN^2-1\right)}}\right)\left[\left(1+\dfrac{4\msqq}{\hat{s}}\right)\beta+\left(\dfrac{4\msqq}{\hat{s}}-\dfrac{8\msqqu}{\hat{s}^2}\right)\ln\left(\dfrac{1-\beta}{1+\beta}\right)\right];  \\
\hat{\sigma}^{(0)}_{3}&=\dfrac{\pi\als^2}{\hat{s}}\left({\dfrac{\left(n_f-1\right)\SUNN}{\SUNN^2-1}}\right)\left[\left(\dfrac{1}{6}+\dfrac{16\msqq}{3\hat{s}}\right)\beta+\left(\dfrac{2\msqq}{\hat{s}}+\dfrac{4\msqqu}{\hat{s}^2}\right)\ln\left(\dfrac{1-\beta}{1+\beta}\right)\right].
\end{align}
At threshold this becomes:
\begin{align}
\hat{\sigma}^{(0,\myth)}_{1}&=\pi\als^2\left({\dfrac{n_f-1}{\SUNN\left(\SUNN^2-1\right)}}\right)\left[\dfrac{\beta}{4\msqq}\right]; \\
\hat{\sigma}^{(0,\myth)}_{2}&=\pi\als^2\left({\dfrac{\left(n_f-1\right)\left(\SUNN^2-4\right)}{2\SUNN\left(\SUNN^2-1\right)}}\right)\left[\dfrac{\beta}{4\msqq}\right];\\
\hat{\sigma}^{(0,\myth)}_{3}&=\pi\als^2\left({\dfrac{\left(n_f-1\right)\SUNN}{\SUNN^2-1}}\right)\left[\dfrac{\beta^3}{24\msqq}\right].
\end{align}
As expected, this channel shows no difference with respect to the MSSM, as neither gluinos nor sgluons play a role at the lowest order. Since the singlet and symmetric octet are in an $s$-wave state, they can be resummed at NNLL accuracy.

The squark-squark production initiated by two quarks of any flavour can be decomposed as:
\begin{equation*}
q_i(a_1)q_j(a_2)\rightarrow \sq(a_3)\sq(a_4)~\text{(+h.c.)}\qquad \mathbf{\SUNN} \otimes \mathbf{\SUNN} = \dfrac{\mathbf{\SUNN}\left(\mathbf{\SUNN-1}\right)}{\mathbf{2}} \oplus \dfrac{\mathbf{\SUNN}\left(\mathbf{\SUNN+1}\right)}{\mathbf{2}}
\end{equation*}
\begin{align*}
c_{1} &=\dfrac{1}{2}\left(\delta_{a_1a_4}\delta_{a_2a_3}-\delta_{a_1a_3}\delta_{a_2a_4}\right); &\qquad C_2\left(R_{1}\right)&= \dfrac{\left(\SUNN+1\right)\left(\SUNN-2\right)}{\SUNN};\\
c_{2} &=\dfrac{1}{2}\left(\delta_{a_1a_4}\delta_{a_2a_3}+\delta_{a_1a_3}\delta_{a_2a_4}\right); &\qquad C_2\left(R_{2}\right)&= \dfrac{\left(\SUNN-1\right)\left(\SUNN+2\right)}{\SUNN}.
\end{align*}
They correspond to $R_1=\bar{\mathbf{{3}}}$ and $R_2=\mathbf{6}$ for $\mathrm{SU(3)}$. The LO cross section~[Eq.(3.5) of~\cite{Diessner:2017ske}] can then be split into:
\begin{align}
\hat{\sigma}^{(0)}_{1}&=\dfrac{\pi\alssusy^2}{\hat{s}}\dfrac{(\SUNN-1)(\SUNN+1)^2}{4\SUNN^3}\left[-2\beta-\left(1+\dfrac{2m^2_-}{\hat{s}}\right)L_1\right];\\
\hat{\sigma}^{(0)}_{2}&=\dfrac{\pi\alssusy^2}{\hat{s}}\dfrac{(\SUNN+1)(\SUNN-1)^2}{4\SUNN^3}\left[-2\beta-\left(1+\dfrac{2m^2_-}{\hat{s}}\right)L_1\right].
\end{align}
If one expands near threshold, $\beta\rightarrow0$, one obtains
\begin{align}
\hat{\sigma}^{(0,\myth)}_{1}&=\pi\alssusy^2\dfrac{(\SUNN-1)(\SUNN+1)^2}{4\SUNN^3}\left[\beta^3\dfrac{2\msqq}{3\left(\mglq+\msqq\right)^2}\right]; \\
\hat{\sigma}^{(0,\myth)}_{2}&=\pi\alssusy^2\dfrac{(\SUNN+1)(\SUNN-1)^2}{4\SUNN^3}\left[\beta^3\dfrac{2\msqq}{3\left(\mglq+\msqq\right)^2}\right].
\end{align}
The production of squarks is power-suppressed in the MRSSM due to the absence of the gluino mass in the numerator. This constitutes a big difference to the MSSM as here, we can only resum up to NLL accuracy.

\section{Expressions for the $C$-coefficients}
\label{appendix: ccoeffs}
In this appendix, we present the results for the one-loop matching coefficients $C^{(1)}_{gg\to \sq\sq^*, I}$ of Eq.~\eqref{eq:res_exp} for squark-antisquark production in the gluon channel\footnote{Recall that the matching coefficients for squark-antisquark production in the quark-antiquark channel as well as the matching coefficients for squark-squark production, both of which occur in a $p$-wave state only, are zero, $C^{(1)}_{qq\to \sq\sq, I} = C^{(1)}_{q\bar q\to \sq\sq^*, I} = 0$, as we take these channels into account only up to NLL accuracy.}. The matching coefficient can be split into a part corresponding to the one-loop Coulomb corrections and the one-loop hard-matching coefficient:
\begin{equation}
	C^{(1)}_{gg\to \sq\sq^*, I} = \mathcal{C}^{\text{Coul},(1)}_{gg\to \sq\sq^*, I}(N, \{m^2\}, \mu^2) + \mathcal{C}^{(1)}_{gg\to \sq\sq^*, I}(\{m^2\}, \mu^2)\,.
\end{equation}
The expressions for the Coulomb contributions $\mathcal{C}^{\text{Coul},(1)}_{gg\to \sq\sq^*, I}$ can be found in App.~A of \cite{Beenakker:2014sma}. In the following, we thus only show the results for the hard-matching coefficients $\mathcal{C}^{(1)}_{gg\to \sq\sq^*, I}$.

The results are presented by introducing suitable abbreviations and carrying out
simplifications. We did not change the arguments of the logarithms or
polylogarithms resulting from the \texttt{FeynArts/FormCalc}
computation in order to avoid any problems from branch cuts/imaginary parts.

\subsection*{Singlet}
In the MSSM, the singlet part of the hard-matching coefficient can be
written as
\begin{align}
\mathcal{C}^{\mathrm{(1)}}_{\mathrm{MSSM},\mathbf{1}} &= \operatorname{Re}\Bigg\{{-4} + \dfrac{11 \pi ^2}{12}+6 \mygamma ^2+6 \mygamma  \log \left(\dfrac{\mu^2_F}{4\msqq}\right)+\dfrac{23}{6}\log \left(\dfrac{\mu^2_R}{\mu^2_F}\right)\nonumber\\
&\qquad\quad\, {}+\dfrac{2}{3}\left(1-\dfrac{\mglqu}{\msqqu}\right) \log\left(\dfrac{\mglq-\msqq}{\mglq+\msqq}\right) - \dfrac{4 \mglq}{3 \msqq}\nonumber\\
&\qquad\quad\, {}-\dfrac{3 \mglq}{2 \msqq} \log ^2\left(\dfrac{2 \msq\left(\sqrt{\msqq-\mglq}-\msq\right)}{\mglq}+1\right)\nonumber\\
&\qquad\quad\, {}+ \dfrac{3 \left(\mglq+\msqq\right) }{2 \msqq}\left[\polylog\left(-\dfrac{\msqq}{\mglq}\right)-\polylog\left(\dfrac{\msqq}{\mglq}\right)\right]\Bigg\}.
\end{align}
In the MRSSM, the following additional singlet terms arise,
\begin{align}
\mathcal{C}^{\mathrm{(1)}}_{\mathrm{MRSSM},\mathbf{1}}-\mathcal{C}^{\mathrm{(1)}}_{\mathrm{MSSM},\mathbf{1}}&= \operatorname{Re}\Bigg\{ \frac{2}{3} \mglq \left(\frac{\pi^2}{\mOs^2}+\frac{4}{\msqq}\right) \nonumber\\
&\qquad\quad\, {} + \left[3 A_1\left(\msqq,\mglq,\mOs^2\right)+A_4\left(\msqq,\mglq,\mOs^2\right)\right] \mOs^2\nonumber\\
&\qquad\quad\, {} + 8 A_4\left(\msqq,\mglq,\mOs^2\right) \msqq+6 A_5\left(\msqq,\mglq,\mOs^2\right)\nonumber\\
&\qquad\quad\, {} - 8 A_7\left(\msqq,\mglq,\mOs^2\right)\Bigg\}.
\end{align}

\subsection*{Symmetric octet}
The MSSM octet part of the hard-matching coefficient can be written as
\begin{align}
\mathcal{C}^{\mathrm{(1)}}_{\mathrm{MSSM},\mathbf{8_S}}&=\operatorname{Re}\Bigg\{2+3\mygamma+\dfrac{31 \pi ^2}{24}+6 \mygamma ^2+6 \mygamma  \log \left(\dfrac{\mu^2_F}{4\msqq}\right)+\dfrac{23}{6}\log \left(\dfrac{\mu^2_R}{\mu^2_F}\right)\nonumber\\
&\qquad\quad\, {} - \dfrac{4 \mglq}{3 \msqq}-\dfrac{3 \mglq}{4 \msqq} \log ^2\left(\dfrac{2 \msq\left(\sqrt{\msqq-\mglq}-\msq\right)}{\mglq}+1\right)\nonumber\\
&\qquad\quad\, {} + \dfrac{3 \mglq }{2 \msqq}\left[\polylog\left(-\dfrac{\msqq}{\mglq}\right)-\polylog\left(\dfrac{\msqq}{\mglq}\right)\right]+\dfrac{2}{3}\left(1-\dfrac{\mglqu}{\msqqu}\right) \log\left(\dfrac{\mglq-\msqq}{\mglq+\msqq}\right)\Bigg\}.
\end{align}
The additional MRSSM octet terms are
\begin{align}
\mathcal{C}^{\mathrm{(1)}}_{\mathrm{MRSSM},\mathbf{8_S}}-\mathcal{C}^{\mathrm{(1)}}_{\mathrm{MSSM},\mathbf{8_S}} &= \operatorname{Re}\Bigg\{\left[3 A_1\left(\msqq,\mglq,\mOs^2\right)+A_4\left(\msqq,\mglq,\mOs^2\right)\right] \left(\mOs^2-\msqq\right)\nonumber\\
&\qquad\quad\, {} + 3 A_5\left(\msqq,\mglq,\mOs^2\right)-8 A_7\left(\msqq,\mglq,\mOs^2\right)\nonumber\\
&\qquad\quad\, {} +\mglq \left(\frac{8}{3\msqq}-\frac{\pi^2}{12\mOs^2}\right)\Bigg\}.
\end{align}
Here, the following abbreviations for combinations of dilogarithms and
logarithms have been used:
\begin{align}
A_1\left(\msqq,\mglq,\mOs^2\right)&=\dfrac{\mglq\mOs^2}{2\msqq(\mOs^4 - 3\mOs^2\msqq + 2\msqqu)}\Bigg\{L_{14}^2-L_{13}^2 \nonumber\\
&\qquad  {} + 2\Bigg[{-\polylog\left(\dfrac{2(\mOs^2 - \msqq)}{\mOs(\mOs + \tilde{r}^-_{\Os})}\right)} + \polylog\left(\dfrac{2(\mOs^2 - \msqq)}{\mOs^2 - 2\msqq + \mOs \tilde{r}^-_{\Os}}\right) \nonumber \\
&\qquad\qquad {}- \polylog\left( \dfrac{2(\mOs^2 - 2\msqq)}{\mOs^2 - 2\msqq - r^{+,4}_{\Os}}\right)+\polylog\left( \dfrac{2(\mOs^2 - \msqq)}{\mOs^2 - 2\msqq - r^{+,4}_{\Os}}\right) \nonumber\\
&\qquad\qquad {}- \polylog\left( \dfrac{2(\mOs^2 - 2\msqq)}{\mOs^2 - 2\msqq + r^{+,4}_{\Os}}\right)+\polylog\left( \dfrac{2(\mOs^2 - \msqq)}{\mOs^2 - 2\msqq + r^{+,4}_{\Os}}\right) \Bigg] \Bigg\};\\
A_4\left(\msqq,\mglq,\mOs^2\right)&=\dfrac{\mglq}{6\mOs^2\left(\mOs^2-\msqq\right)}\Bigg\{L_{8}^2-L_{9}^2+L_{15}^2 \nonumber\\
&\qquad {} + 2\Bigg[{\polylog\left(\dfrac{2(\mOs^2 - \msqq)}{\mOs(\mOs + \tilde{r}^-_{\Os})}\right)} + \polylog\left(\dfrac{2(\mOs^2 - 2\msqq)}{\mOs^2 - 2\msqq - \mOs \tilde{r}^-_{\Os}}\right) \nonumber \\
&\qquad\qquad {}+\polylog\left( \dfrac{2(\mOs^2 - 2\msqq)}{\mOs^2 - 2\msqq + \mOs \tilde{r}^-_{\Os}}\right)-\polylog\left( \dfrac{2(\mOs^2 - \msqq)}{\mOs^2 - 2\msqq + \mOs \tilde{r}^-_{\Os}}\right) \nonumber \\
&\qquad\qquad {}+ \polylog\left( \dfrac{2(\mOs^2 - 2\msqq)}{\mOs^2 - r^{+,4}_{\Os}}\right)-\polylog\left( \dfrac{2(\mOs^2 - \msqq)}{\mOs^2 -2\msqq- r^{+,4}_{\Os}}\right) \Bigg] \Bigg\};\\
A_5\left(\msqq,\mglq,\mOs^2\right)&=\dfrac{\mglq \mOs^2}{4\mOs^2\msqq - 8\msqqu}L_{12}^2; \\
A_7\left(\msqq,\mglq,\mOs^2\right)&=\dfrac{\mglq}{6\msqqu\; \tilde{r}^-_{\Os}}\Bigg[\mOs\left(\mOs^2-2\msqq\right)\left(\log\left(\dfrac{2\msq}{\mOs}\right)-L_{7}\right)\nonumber \\
&\qquad\qquad\quad~ {} + \tilde{r}^-_{\Os}\;r^{+,4}_{\Os}\left(\log\left(\dfrac{\mOs}{2\msq}\right)+L_{10}\right)\Bigg].
\end{align}
Further abbreviations for square roots and logarithms are
\begin{align}
r^-_{\Os}&= \sqrt{\msqq-\mOs^2}, \qquad &&r^-_{\gl}= \sqrt{\msqq-\mglq},\\
\tilde{r}^-_{\Os}&= \sqrt{\mOs^2-4\msqq}, \qquad &&r^{+,4}_{\Os}= \sqrt{\mOs^4+4\msqqu},
\end{align}
\begin{align}
L_{7} &= \log\left(\dfrac{\mOs + \tilde{r}^-_{\Os}}{\mOs}\right), \\
L_{8} &= \log\left(\dfrac{r^{+,4}_{\Os}-\mOs^2}{r^{+,4}_{\Os}+\mOs^2}\right), \\
L_{9} &= \log\left(\dfrac{r^{+,4}_{\Os}+2\msqq-\mOs^2}{r^{+,4}_{\Os}+\mOs^2}\right), \\
L_{10} &= \log\left(\dfrac{r^{+,4}_{\Os}+2\msqq+\mOs^2}{\mOs^2}\right), \\
L_{12} &= \log\left(\dfrac{\mOs^2 - 2\msqq + 2\msq r^-_{\Os}}{\mOs^2}\right), \\
L_{13} &= \log\left(\mOs(-\mOs + \tilde{r}^-_{\Os})\right)- \log\left(\mOs(\mOs + \tilde{r}^-_{\Os})\right), \\
L_{14} &=  \log\left(\mOs^2 - 2\msqq + \mOs \tilde{r}^-_{\Os}\right)-\log\left(\mOs(-\mOs + \tilde{r}^-_{\Os})\right) , \\
L_{15} &=  \log\left(-\mOs^2 + 2\msqq + \mOs \tilde{r}^-_{\Os}\right)-\log\left(\mOs(\mOs + \tilde{r}^-_{\Os})\right) .
\end{align}

\printbibliography[title={References}]
\addcontentsline{toc}{section}{References}

\end{document}